\documentclass{article}
\usepackage{iclr2027_conference,times}

\usepackage{amsmath,amsfonts,bm}

\def\eqref#1{equation~\ref{#1}}

\def\1{\bm{1}}

\DeclareMathAlphabet{\mathsfit}{\encodingdefault}{\sfdefault}{m}{sl}
\SetMathAlphabet{\mathsfit}{bold}{\encodingdefault}{\sfdefault}{bx}{n}

\usepackage{amsmath,amssymb}
\usepackage{booktabs}
\usepackage{graphicx}
\usepackage{xcolor}
\usepackage{tikz}
\usepackage{hyperref}
\usepackage{url}
\usepackage{xspace}
\usetikzlibrary{arrows.meta,calc,fit,patterns,positioning}
\usepackage{multirow}
\usepackage{algorithm}
\usepackage{algorithmic}
\usepackage{subcaption} 
\usepackage{adjustbox}
\usepackage{siunitx}
\usepackage{tabularx}

\usepackage{booktabs}
\usepackage[dvipsnames, svgnames]{xcolor}
\usepackage{amssymb}
\usepackage{natbib}
\usepackage[table]{xcolor}
\usepackage{listings}
\usepackage{tcolorbox}
\tcbuselibrary{listings}

\definecolor{easyblue}{HTML}{177E89}
\definecolor{easybluefill}{HTML}{DCEFF1}
\definecolor{baselineorange}{HTML}{C86B32}
\definecolor{baselinefill}{HTML}{F7E5D8}
\definecolor{ink}{HTML}{263238}
\definecolor{midgray}{HTML}{6B7280}
\definecolor{lightgray}{HTML}{E5E7EB}
\definecolor{palegray}{HTML}{F5F6F7}

\title{Coarse-to-Fine Macro Placement via Evolutionary Search and Critical Macro Tuning}

\author{Biao Liu, Zhiping Jin, Kaixuan Sun, Zengrui Lu, Qingquan Zhang, Bo Yuan\thanks{Corresponding author.}\\
Department of Computer Science and Engineering\\
Southern University of Science and Technology\\
Shenzhen, China\\
}

\iclrfinalcopy

\begin{document}

\maketitle

\begin{abstract}
Macro placement is a critical stage in chip physical design that substantially affects downstream implementation quality. Recent search-based methods improve existing layouts through partial reconstruction, but quality-biased or spatially restricted macro selection can limit the diversity of reconstruction proposals, potentially hindering escape from local optima. Moreover, coarse-grid representations restrict placement precision. To address these challenges, we propose C2FPlace, a \textbf{C}oarse-to-\textbf{F}ine macro \textbf{Place}ment framework that integrates population-based evolutionary search with fine-grained refinement. During coarse-grained optimization, tournament selection chooses promising parents from randomly sampled groups of layouts, and stochastic partial rip-up and re-place generates offspring by sampling macro subsets across the entire layout. A two-phase schedule samples reconstruction ratios from a higher range early in the search and a lower range later, supporting broad exploration followed by more conservative refinement. During fine-grained optimization, critical macro tuning enables positional adjustments beyond the coarse grid to obtain additional half-perimeter wirelength (HPWL) reduction. Experiments on the ISPD2005 benchmark show that C2FPlace reduces HPWL by 17.82\% over EGPlace and 17.86\% over RollPlace on average. On the ICCAD2025 benchmark, C2FPlace achieves the best average ranking among the compared methods under the evaluated power, performance, and area (PPA) metrics. Our codes are available in \href{https://github.com/lxxxxb/C2FPlace}{https://github.com/lxxxxb/C2FPlace}.
\end{abstract}

\section{Introduction}
\label{sec:introduction}

Macro placement is a critical stage in chip physical design that significantly affects downstream design closure. In modern VLSI systems, high-quality macro placement facilitates subsequent standard cell placement, clock tree synthesis, and detailed routing. Conversely, poor macro placement can degrade implementation quality even after extensive downstream optimization~\cite{kang1994fuzzy,macmillen2000industrial}. The macro placement problem is NP-hard, with tightly coupled decisions arising from macro dimensions, connectivity, and non-overlap constraints. Analytical placers address placement through continuous optimization and gradient-based numerical techniques~\cite{chengRePlAceAdvancingSolution2019,lin2019dreamplace,liaoDREAMPlace40Timingdriven2022}. Although these methods offer strong scalability, their solutions can depend on initialization and subsequent legalization, which may alter the optimized placement objectives.

Recent reinforcement learning (RL) and black-box optimization (BBO) methods have shown promising results for macro placement~\cite{graphplace,lai2022maskplace,lai2023chipformer,shi2023macro,efficientplace,maskregulate,liegplace,zhou2025rollplace,geng2025lamplace}. Beyond constructing layouts sequentially, recent search-based approaches improve existing solutions through partial reconstruction while retaining the positions of unselected macros, as shown in Table~\ref{tab:macro_ordering}. EGPlace~\cite{liegplace} maintains a population of layouts and uses fitness-based probabilities to select parents for evolution. RollPlace~\cite{zhou2025rollplace} organizes candidate layouts in a Monte Carlo search tree and uses the Upper Confidence Bound applied to Trees (UCT) criterion for candidate selection. These methods demonstrate the value of maintaining multiple candidates and reusing existing placement structures.

Despite these advances, generating sufficiently diverse reconstruction proposals remains an important challenge. In EGPlace, parent selection follows a softmax distribution over fitness values, while macro selection is guided by scores derived from wirelength, congestion, and overlap. Although both selections are stochastic, their preference for current quality estimates may concentrate search effort on similar candidates and macro subsets. RollPlace samples two coordinates to define a rectangular region and repositions all macros within that region. Such region-based selection couples macro inclusion to spatial proximity, restricting the subsets that can be reconstructed together in a single update. These selection biases may limit exploration of alternative macro combinations and make it harder to escape local optima. This motivates combining quality-aware parent selection with macro-subset sampling that is not tied to current macro scores or a single rectangular region.

The extent of reconstruction introduces a further exploration--exploitation trade-off. Repositioning a large subset of macros permits broader coordinated changes, but also exposes more established placement relationships to modification. Repositioning a small subset retains more of the current structure, but offers fewer opportunities for substantial rearrangement. A fixed reconstruction ratio, as used in the reported EGPlace experiments, does not explicitly reflect these differing needs across search stages. Varying the reconstruction scale over time offers a way to encourage broad exploration early in the search and preserve more of the evolved layout during later refinement.

A second challenge arises from placement granularity. Many recent methods discretize the placement canvas into a coarse grid, such as \(224\times224\), to make the search space tractable. While this representation reduces optimization complexity, it also restricts available macro coordinates. Even after effective coarse-grid search, finer positional adjustments may further improve wirelength, particularly for macros whose pins determine net bounding boxes. Combining coarse-grained exploration with targeted fine-grained refinement can exploit these remaining opportunities without repeating global search at a finer resolution.

\begin{table}[t]
\centering
\caption{Comparison of macro placement methods and their search mechanisms.}
\label{tab:macro_ordering}
\begingroup
\definecolor{c2fBaselineRow}{RGB}{237,244,250}
\definecolor{c2fOursRow}{RGB}{252,239,222}
\footnotesize
\setlength{\tabcolsep}{2pt}
\newcommand{\cTwoFMethod}[2]{#1~{\footnotesize\citep{#2}}}
\begin{tabular*}{\textwidth}{@{\extracolsep{\fill}}lcccc@{}}
\toprule
\textbf{Method} & \textbf{Parent selection} & \textbf{Macro selection} & \textbf{Update scale} & \textbf{Granularity} \\
\midrule
\cTwoFMethod{MaskPlace}{lai2022maskplace} & N/A & N/A & N/A & Coarse grid \\
\cTwoFMethod{ChiPFormer}{lai2023chipformer} & N/A & N/A & N/A & Coarse grid \\
\cTwoFMethod{WireMask-BBO}{shi2023macro} & 1+1 EA & N/A & All & Coarse grid \\
\cTwoFMethod{EfficientPlace}{efficientplace} & UCT & N/A & N/A & Coarse grid \\
\cTwoFMethod{MaskRegulate}{maskregulate} & Fixed & N/A & All & Coarse grid \\
\midrule
\rowcolor{c2fBaselineRow}
\cTwoFMethod{EGPlace}{liegplace} & Fitness-based & Score-guided & Fixed ratio & Coarse grid \\
\rowcolor{c2fBaselineRow}
\cTwoFMethod{RollPlace}{zhou2025rollplace} & UCT & Rectangle & Region-based & Coarse grid \\
\midrule
\rowcolor{c2fOursRow}
\textbf{C2FPlace (ours)} & Tournament & Random subset & Two-phase ratio & \textbf{Coarse-to-fine} \\
\bottomrule
\end{tabular*}
\endgroup
\end{table}

To address these challenges, we propose C2FPlace, a coarse-to-fine macro placement framework centered on population-based evolutionary search. C2FPlace first generates multiple initial layouts through repeated WireMask-guided~\cite{lai2022maskplace,shi2023macro} greedy placement. During coarse-grained optimization, tournament selection randomly samples a group of individuals and chooses the best among them as the parent. This provides quality-aware selection based on relative ranking within each sampled group, without requiring a softmax transformation of fitness values. Offspring are generated through stochastic partial rip-up-and-replace: a reconstruction ratio is sampled, and the corresponding subset of macros is randomly selected across the entire layout for repositioning, while unselected macros retain their positions. A two-phase schedule draws ratios from a higher range during the initial search phase and a lower range during the subsequent phase. Together, random parent-group sampling, random macro-subset selection, and varying reconstruction ratios provide multiple sources of search variation, while tournament selection and elitist replacement favor promising layouts. The schedule shifts the search from broad reconstruction toward more conservative refinement.

Following coarse-grained search, C2FPlace performs fine-grained critical macro tuning to further reduce half-perimeter wirelength (HPWL). This stage selectively adjusts critical macros beyond the coarse-grid coordinates while preserving the global arrangement established by evolutionary search. Coarse-grained optimization thus provides the main search capability, and fine-grained tuning complements it with additional positional refinement. Our main contributions are summarized as follows:
\begin{itemize}
    \item We propose a population-based evolutionary search strategy that combines tournament selection with stochastic partial rip-up and re-place. Random macro-subset sampling across the layout promotes diverse reconstruction proposals, while a two-phase reconstruction-ratio schedule supports broad exploration followed by more conservative refinement.
    \item We propose a coarse-to-fine framework that integrates population-based search on a coarse placement grid with fine-grained critical macro tuning, enabling additional HPWL reduction through positional adjustments beyond the grid.
    \item Experiments on the ISPD2005 benchmark C2FPlace reduces HPWL by 17.82\% over EGPlace and 17.86\% over RollPlace on average. On the ICCAD2025 benchmark, C2FPlace achieves the best average ranking among the compared methods under the evaluated PPA metrics.
\end{itemize}

\section{Related Work}
\label{sec:related_work}

\subsection{Learning-based Methods}
Learning-based methods, particularly reinforcement learning (RL), learn policies for constructing or refining macro placements. GraphPlace~\cite{graphplace} formulates macro placement as a sequential Markov decision process, while DeepPR~\cite{deeppr} and PRNet~\cite{prnet} incorporate feedback from standard-cell placement and routing. MaskPlace~\cite{lai2022maskplace} uses pixel-wise masks to represent placement constraints and wirelength information, and GoodFloorPlan~\cite{goodfloorplan} combines graph convolutional networks with sequence-pair representations. To improve learning efficiency, ChiPFormer~\cite{lai2023chipformer} adopts offline RL with placement-data pretraining, whereas EfficientPlace~\cite{efficientplace} integrates Monte Carlo Tree Search with RL. More recently, MaskRegulate~\cite{maskregulate} focuses on iterative refinement and incorporates regularity priors into its state representation and reward design to improve placement quality and downstream PPA.

\subsection{Optimization-based Methods}
Optimization-based approaches include partition-based, analytical, and meta-heuristic methods. Partition-based methods~\cite{breuerClassMincutPlacement1988,agnihotriRecursiveBisectionPlacement2005,royMincutFloorplacement2006} recursively divide the placement region and netlist, while Hier-RTLMP~\cite{kahng2024hier} applies hierarchical decomposition to complex designs. Analytical methods use quadratic~\cite{spindler2007fast,spindler2008kraftwerk2,linPOLAR30Ultrafast2015,liUTPlaceFRoutabilitydrivenFPGA2016,chengNetSeparationOrientedPrinted2022} or nonlinear formulations~\cite{chen2008ntuplace3,kahngImplementationExtensibilityAnalytic2005,hsuNTUplace4hNovelRoutabilityDriven2014}. Electrostatic approaches include ePlace-MS~\cite{lu2015eplace}, ePlace~\cite{luEPlaceElectrostaticsBasedPlacement2015}, and RePlAce~\cite{chengRePlAceAdvancingSolution2019}; other scalable frameworks include PeF~\cite{li2022pef}, DREAMPlace~\cite{lin2019dreamplace}, and Xplace~\cite{liu2023xplace}. These methods generally require subsequent legalization to obtain feasible layouts. Meta-heuristic approaches include evolutionary search~\cite{cohoon1987genetic} and simulated annealing~\cite{shunmugathammalNovelBtreeCrossoverBased2020,kahngRTLMPPracticalHumanQuality2022}. AutoDMP~\cite{agnesinaAutoDMPAutomatedDREAMPlacebased2023} uses Bayesian optimization to tune DREAMPlace~\cite{lin2019dreamplace}, while WireMask-BBO~\cite{shi2023macro} combines black-box search with WireMask-guided greedy placement. EGPlace~\cite{liegplace} evolves a population through score-guided macro repositioning, and RollPlace~\cite{zhou2025rollplace} combines Monte Carlo tree search with order exchanges and repeated rollouts. Building on partial reconstruction, C2FPlace combines tournament selection with scheduled random-subset repositioning, followed by fine-grained critical macro tuning.

\section{Preliminary}
\subsection{Problem Formulation}
The input of a macro placement problem is a netlist, which contains wire nets and module information, such as macro sizes and pin offsets. A wire net connects a group of pins from different modules. A placement solution is represented as $P=\{(x_i, y_i)\}_{i=1}^{N}$, where $(x_i,y_i)$ denotes the position of the $i$-th macro and $N$ is the total number of macros. Since exact routing evaluation is computationally expensive, heuristic metrics such as half-perimeter wirelength (HPWL) are commonly used to guide macro placement. HPWL provides an efficient wirelength estimation by converting routing optimization into a geometric calculation \cite{graphplace,gareyRectilinearSteinerTree1977}. This paper focuses on minimizing HPWL under non-overlap constraints, formulated as:
\begin{equation}
\min_{P}\ \text{HPWL}(P),\quad \text{s.t.}\ \text{Overlap}(x,y,w,h)=0
\end{equation}
where $x$ and $y$ represent macro positions, and $w$ and $h$ represent macro width and height, respectively.
HPWL is defined as the sum of the half-perimeters of the bounding boxes for all wire nets in the netlist, which is formulated as: 
\begin{equation}
\begin{split}
\operatorname{HPWL} 
= \sum_{e_i \in E} \Bigg[
& \left( \max_{p \in e_i} x_p - \min_{p \in e_i} x_p \right) + \left( \max_{p \in e_i} y_p - \min_{p \in e_i} y_p \right)
\Bigg],
\end{split}
\label{eq:hpwl}
\end{equation}
where \( E \) is the set of all wire nets, and \( p \) denotes a pin in net \( e_i \) with coordinates \( (x_p, y_p) \).

\subsection{WireMask-Based Placement Operator}
\label{sec:wiremask}
C2FPlace employs a WireMask-based placement operator~\cite{lai2022maskplace,shi2023macro} as the basic placement primitive. 
Given a macro to be placed, the operator first identifies legal locations satisfying boundary and non-overlap constraints. 
For each legal location, WireMask estimates the induced wirelength by computing the HPWL of nets connected to the macro, and the location with the minimum estimated wirelength is selected.
Specifically, the placement region is discretized into a $G \times G$ grid. 
For a macro $m$, the legality mask $\mathbf{L}_m$ indicates whether each grid location satisfies placement constraints, while the WireMask $\mathbf{W}_m$ records the estimated HPWL cost of placing $m$ at each legal location. 
The selected position is obtained by:
\begin{equation}
(i_m,j_m)=
\arg\min_{(i,j)}
\{\mathbf{W}_m(i,j)\mid \mathbf{L}_m(i,j)=1\}.
\end{equation}
Here, $(i_m, j_m)$ denotes a coarse-grained grid index, which must be mapped to physical coordinates. Fine-grained tuning evaluates the same wirelength cost on its fine-grained candidate set. This placement operator is consistently used in initialization, evolutionary search, and critical macro tuning, providing a unified optimization primitive throughout C2FPlace.

\begin{figure*}[t]
  \centering 
  \includegraphics[width=1.0\textwidth]{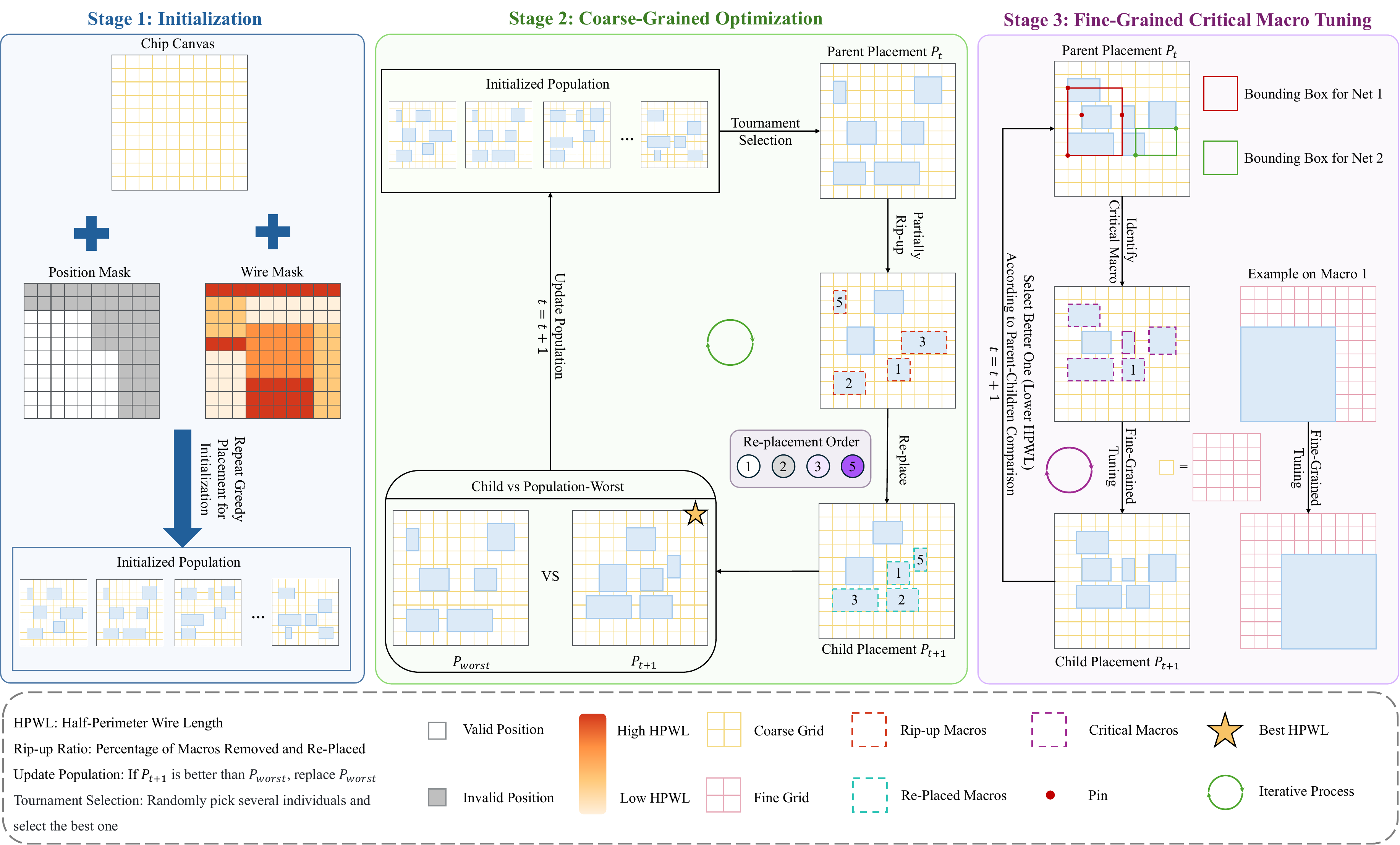}
  \caption{Overview of C2FPlace. The framework initializes a population of diverse greedy placements, evolves them through stochastic evolutionary search with partial rip-up and re-place, and performs critical macro tuning for further HPWL reduction.} 
  \label{fig:myimage} 
\end{figure*}
\section{Method}
Figure~\ref{fig:myimage} illustrates the overall pipeline of C2FPlace. Starting from a population of diverse WireMask-guided initial placements, C2FPlace performs evolutionary optimization in two stages. The coarse-grained stage uses tournament selection and a staged exploration-to-exploitation reconstruction strategy to explore diverse solutions while preserving high-quality structures. The resulting layouts are further refined by fine-grained critical macro tuning to correct discretization errors from coarse-grid placement. WireMask-guided greedy placement serves as the common placement operator throughout the search.

\subsection{Initial Population Generation}
The optimization starts from a diverse population generated by the WireMask-based placement operator. Instead of relying on a single individual initialization, C2FPlace independently constructs 20 initial placement solutions through greedy WireMask-guided macro placement.

During initialization, macros are placed according to a deterministic ordering heuristic that prioritizes larger and more strongly connected macros. This strategy alleviates early resource contention and produces stable placement configurations. Crucially, when multiple locations offer equivalent optimality, the algorithm selects randomly among them. By repeatedly applying the greedy placement procedure, C2FPlace obtains a diverse initial population, providing multiple promising starting points for subsequent evolutionary optimization.
\subsection{Coarse-Grained Evolutionary Search}
C2FPlace optimizes the initial population through a steady-state evolutionary search. At each iteration, five individuals are randomly sampled for tournament selection, and the one with the lowest HPWL is chosen as the parent. The parent undergoes partial rip-up and re-place to generate an offspring, which is evaluated and incorporated into the population through elitist update, as described in Section~\ref{sec:elitist}.
\subsubsection{Mutation Strategy}
Inspired by evolutionary optimization, C2FPlace generates new candidate placements through stochastic mutation rather than complete reconstruction. Our mutation strategy consists of three complementary components: partial rip-up and re-place, staged exploration-to-exploitation scheduling, and heuristic-guided reinsertion.
\paragraph{Partial Rip-up and Re-place}
Instead of reconstructing the entire placement, C2FPlace performs optimization by modifying only a subset of macros in each iteration. Specifically, a subset of macros is randomly selected and removed from the current placement, while all remaining macros remain fixed. Compared with reconstruction, this strategy reduces optimization cost while preserving high-quality placement structures accumulated during previous iterations, thereby providing a better balance between search diversity and solution stability.
\paragraph{Exploration-to-Exploitation}
To balance these competing objectives,C2FPlace uses a two-phase schedule for exploration-to-exploitation schedule. At the first phase, the rip-up ratio is sampled from a high-valued interval, providing stronger perturbation for global exploration.  At the second phase, The sampling interval shifts toward lower values, enabling a transition toward exploitation.
\paragraph{Heuristic-Guided Reinsertion}
After determining the rip-up ratio, the selected macros are removed from the current placement and reinserted one by one using the WireMask-based placement operator described in Section~\ref{sec:wiremask}. C2FPlace follows a topology-aware heuristic ordering that prioritizes macros with higher connectivity and larger area.

\subsubsection{Population Update}
\label{sec:elitist}

After mutation, the generated offspring is evaluated based on its HPWL. To preserve high-quality solutions while continuously improving the population, C2FPlace adopts an elitist replacement strategy. Specifically, the offspring is compared with the worst individual in the current population and replaces it only when achieving better placement quality. 
Algorithm~\ref{alg:stage2} summarizes the complete coarse-grained evolutionary search procedure.

\subsection{Critical Macro Tuning}
\label{sec:critical_tuning}

Although coarse-grained evolutionary search effectively explores the macro placement space, the obtained solution may still contain suboptimal positions for macros with significant impact on wirelength. This is mainly because coarse-grid representation limits placement flexibility and introduces discretization errors, especially for macros determining net bounding-box boundaries. Therefore, C2FPlace performs critical macro tuning to selectively refine influential macros while preserving the optimized global placement structure.

Different from the global exploration performed in evolutionary search, critical macro tuning focuses on local improvement around high-quality solutions. To balance refinement flexibility and placement precision, C2FPlace adopts a dual-granularity refinement strategy, including structure-preserving coarse-grid displacement and fine-grid offset refinement.

\subsubsection{Critical Macro Identification}

For each net, C2FPlace identifies macros that define the four boundaries of its HPWL bounding box as shown on the right side of Figure~\ref{fig:myimage}. Since displacing these macros can affect the dimensions of the bounding box, they exert the most significant influence on total wirelength.

For net $n$ with pin set $P_n$, the bounding box boundaries are defined as:
\begin{equation}
x_{\min}(n)=\min_{p\in P_n} p.x,\quad
x_{\max}(n)=\max_{p\in P_n} p.x,\quad
y_{\min}(n)=\min_{p\in P_n} p.y,\quad
y_{\max}(n)=\max_{p\in P_n} p.y .
\end{equation}
A macro is classified as an extreme provider if it hosts at least one pin lying on any of these four boundaries. To exclude nets whose bounding box is governed by a single macro, we only consider nets with multiple extreme providers. The critical macro set is then formulated as:
\begin{equation}
\mathcal{M}_{\mathrm{crit}}
=
\bigcup_{\substack{n\\|E_n|\ge2}}
E_n ,
\end{equation}
where $E_n$ represents the set of extreme providers for net $n$.

\subsubsection{Dual-Granularity Refinement}
After identifying the critical macro set, C2FPlace considers two complementary placement strategies for refinement. The first allows larger-scale movement to escape unfavorable coarse-grid positions, while the second focuses on recovering fine-grained precision within the current coarse cell.

\paragraph{Structure-Preserving Coarse-Grid Displacement}
Although coarse-grid placement provides efficient global exploration, a macro may still be trapped at a suboptimal coarse-grid location. Therefore, C2FPlace first performs coarse-grid displacement to explore neighboring coarse locations. Different from unrestricted movement, the displacement is constrained to avoid disturbing surrounding macros. Specifically, the movement from the original location to a candidate location is feasible only when the swept region of the macro does not overlap with other placed macros. Among all feasible candidates, the position with the minimum incremental HPWL is selected. This constraint enables larger-scale local exploration while maintaining the optimized placement structure.

\paragraph{Fine-Grid Offset Refinement}
After coarse-grid displacement, C2FPlace further improves placement precision through fine-grid offset refinement. Instead of changing the coarse-grid location, this strategy searches fine-grained offsets within the current coarse grid cell to compensate for discretization errors. For each critical macro, all feasible offsets within the corresponding coarse cell are evaluated, and the position with the minimum incremental HPWL is selected. This refinement improves placement accuracy while avoiding unnecessary perturbations to the global solution.

\subsubsection{Critical Macro Tuning Procedure}
During each tuning iteration, C2FPlace  first applies coarse-grid displacement to the critical macros. It then re-computes the critical macro set and applies fine-grid offset refinement. The candidate placement is accepted only if its total HPWL is lower than that of the current placement.The WireMask-based placement operator described in Section~\ref{sec:wiremask} evaluates the incremental HPWL of each feasible candidate, and the best position is selected for reinsertion.

To ensure monotonic improvement, C2FPlace adopts an elitist update strategy. The refined placement replaces the current solution only when it achieves lower HPWL. Therefore, the critical macro tuning stage progressively improves local placement quality while preserving the globally optimized placement structure. The pseudocode of Stage~3 is shown in Appendix~\ref{app:detail}.

\section{Experiments}
\subsection{Benchmarks, Baselines and Settings}
For evaluation, we use the ISPD2005 benchmark~\cite{ispd2005} to assess macro placement performance and the ICCAD2025 benchmark~\cite{iccad2025} to evaluate large-scale placement capability and post-placement PPA. We compare C2FPlace with  the
packing-based SP-SA~\cite{murata1996vlsi}, representative analytical placers (NTUPlace3~\cite{chen2008ntuplace3}, RePlAce~\cite{chengRePlAceAdvancingSolution2019}, and DREAMPlace~\cite{lin2019dreamplace}), RL-based approaches (GraphPlace~\cite{graphplace}, DeepPR~\cite{deeppr}, MaskPlace~\cite{lai2022maskplace}, ChiPFormer~\cite{lai2023chipformer}, and EfficientPlace~\cite{efficientplace}), and optimization-based methods (WireMask-EA~\cite{shi2023macro}, EGPlace~\cite{liegplace}, and RollPlace~\cite{zhou2025rollplace}). Detailed settings of C2FPlace are provided in Appendix~\ref{app:hyperparameter}.

\begin{table*}[tbp]
\centering
\caption{HPWL ($\times10^5$) comparison of different macro placement methods on the ISPD2005 benchmark. Lower HPWL values indicate better placement quality. The results of baseline methods are taken from EGPlace and RollPlace. Performance metrics are reported as mean values with standard deviations (mean $\pm$ std), except for the deterministic method NTUPlace3. The best-performing results are highlighted in \textbf{bold}, while the second-best results are marked in \textcolor{brown}{brown}. ``\textit{w/}'' and ``\textit{w/o}'' denote the settings with and without Stage~3 critical macro tuning.}
\label{tab:placement}
\resizebox{\textwidth}{!}{ 
\begin{tabular}{lcccccccc}
\toprule
Method & adaptec1 & adaptec2 & adaptec3 & adaptec4 & bigblue1 & bigblue2 & bigblue3 & bigblue4 \\
\midrule
SP-SA & $18.84 \pm 4.62$ & $117.36 \pm 8.73$ & $115.48 \pm 7.56$ & $120.03 \pm 4.25$ & $5.12 \pm 1.43$ & N/A & $164.70 \pm 19.55$ & N/A \\
NTUPlace3 & 26.62 & 321.17 & 328.44 & 462.93 & 22.85 & N/A & 455.53 & N/A \\
RePlAce & $16.19 \pm 2.10$ & $153.26 \pm 29.01$ & $111.21 \pm 11.69$ & \textcolor{brown}{37.64} \textcolor{brown}{$\pm$} \textcolor{brown}{1.05} & $2.45 \pm 0.06$ & N/A & $119.84 \pm 34.43$ & N/A \\
DREAMPlace & $15.81 \pm 1.64$ & $140.79 \pm 26.73$ & $121.94 \pm 25.05$ & \textbf{37.41} $\pm$ \textbf{0.87} & $2.44 \pm 0.06$ & N/A & $107.19 \pm 29.91$ & $112.95 \pm 4.03$ \\
GraphPlace & $30.10 \pm 2.98$ & $351.71 \pm 38.20$ & $358.18 \pm 13.95$ & $151.42 \pm 9.72$ & $10.58 \pm 1.29$ & N/A & $357.48 \pm 47.83$ & N/A \\
DeepPR & $19.91 \pm 2.13$ & $203.51 \pm 6.27$ & $347.16 \pm 4.32$ & $311.86 \pm 56.74$ & $23.33 \pm 3.65$ & N/A & $430.48 \pm 12.18$ & $433.90 \pm 5.26$ \\
MaskPlace & $7.62 \pm 0.67$ & $75.16 \pm 4.97$ & $100.24 \pm 13.54$ & $87.99 \pm 3.25$ & $3.04 \pm 0.06$ & N/A & $90.04 \pm 4.83$ & $130.15 \pm 10.94$ \\
ChiPFormer & $6.62 \pm 0.05$ & $67.10 \pm 5.46$ & $76.70 \pm 1.15$ & $68.80 \pm 1.59$ & $2.95 \pm 0.04$ & N/A & $72.92 \pm 2.56$ & $94.90 \pm 2.50$ \\
EfficientPlace & $5.94 \pm 0.04$ & $46.79 \pm 1.60$ & $56.35 \pm 0.99$ & $58.47 \pm 1.61$ & $2.14 \pm 0.01$ & $10.48 \pm 0.25$ & $58.38 \pm 0.54$ & $88.01 \pm 2.71$ \\
WireMask-EA & $6.15 \pm 0.05$ & $64.38 \pm 4.43$ & $58.18 \pm 1.04$ & $59.52 \pm 1.71$ & $2.15 \pm 0.01$ & $19.47 \pm 0.17$ & $59.85 \pm 3.39$ & $87.81 \pm 3.30$ \\
EGPlace & \textcolor{brown}{$5.72 \pm 0.01$} & $37.69 \pm 1.08$ & $60.13 \pm 1.83$ & $56.08 \pm 0.43$ & $2.20 \pm 0.01$ & $10.40 \pm 0.34$ & $52.41 \pm 8.16$ & $67.14 \pm 2.42$ \\
RollPlace & 5.93 $\pm$ 0.07 & $34.83 \pm 1.81$ & $54.88 \pm 3.08$ & $64.66 \pm 3.49$ & \textcolor{brown}{2.05 $\pm$ 0.03} & $10.89 \pm 0.22$ & $53.56 \pm 7.31$ & $69.67 \pm 2.70$ \\
C2FPlace \textit{w/o} & \textbf{5.60} $\pm$ \textbf{0.17} & \textcolor{brown}{29.13 $\pm$ 1.91} & \textcolor{brown}{51.19 $\pm$ 2.05} & $43.67 \pm 1.21$ & \textbf{2.04  $\pm$ 0.01} & \textcolor{brown}{7.50 $\pm$ 0.28} & \textcolor{brown}{37.98 $\pm$ 4.32} & \textcolor{brown}{56.29 $\pm$ 1.29} \\
C2FPlace \textit{w/} & \textbf{5.60} $\pm$ \textbf{0.17} & \textbf{29.06} $\pm$ \textbf{1.90} & \textbf{51.06} $\pm$ \textbf{2.05} & $43.39 \pm 1.22$ & \textbf{2.04 $\pm$ 0.01} & \textbf{7.43} $\pm$ \textbf{0.27} & \textbf{37.90 $\pm$ 4.34} & \textbf{56.18} $\pm$ \textbf{1.29} \\
\bottomrule
\end{tabular}
}
\end{table*}

\subsection{ISPD2005 Results}
\subsubsection{Placement Quality Comparison}
We evaluate macro placement quality on the ISPD2005 benchmark containing eight circuits. Following previous works, all macros are placed for \texttt{adaptec1-4}, \texttt{bigblue1}, and \texttt{bigblue3}, while the first 1024 macros are considered for the large-scale \texttt{bigblue2} and \texttt{bigblue4}. The HPWL results are reported in Table~\ref{tab:placement}.

As shown in Table~\ref{tab:placement}, C2FPlace with Stage~3 achieves the lowest HPWL on seven of eight circuits, whereas DREAMPlace performs best on \texttt{adaptec4}. Across the eight benchmarks \texttt{adaptec1-4}, \texttt{bigblue1-4}, C2FPlace reduces HPWL by 2.10\%, 22.90\%, 15.08\%, 22.63\%, 7.27\%, 28.56\%, 27.69\%, and 16.32\% compared with EGPlace, and achieves HPWL reductions of 5.56\%, 16.57\%, 6.96\%, 32.90\%, 0.49\%, 31.77\%, 29.24\%, and 19.36\% relative to RollPlace, respectively. Averaged over all benchmarks, C2FPlace reduces HPWL by 17.82\% against EGPlace and 17.86\% against RollPlace, demonstrating the effectiveness of the proposed coarse-to-fine population-based evolutionary macro placement framework. The visualization for \texttt{adaptec3} is shown in Appendix~\ref{app:visual}. 

Furthermore, we perform ablation experiments to evaluate the contribution of Stage 3. On average, Stage 3 brings a 0.31\% HPWL reduction compared to without Stage 3. These results confirm that Stage 3, which performs fine-grained tuning of critical macros, can further reduce HPWL beyond the previous stages. However, the improvement is negligible in certain scenarios (e.g., \texttt{adaptec1} and \texttt{adaptec4}), where the gain is almost zero. This demonstrates the effectiveness of the proposed coarse-to-fine population-based evolutionary macro placement framework. 

\begin{figure}[tbp]
  \centering
  \begin{subfigure}[b]{0.24\textwidth}
    \centering
    \includegraphics[width=\textwidth]{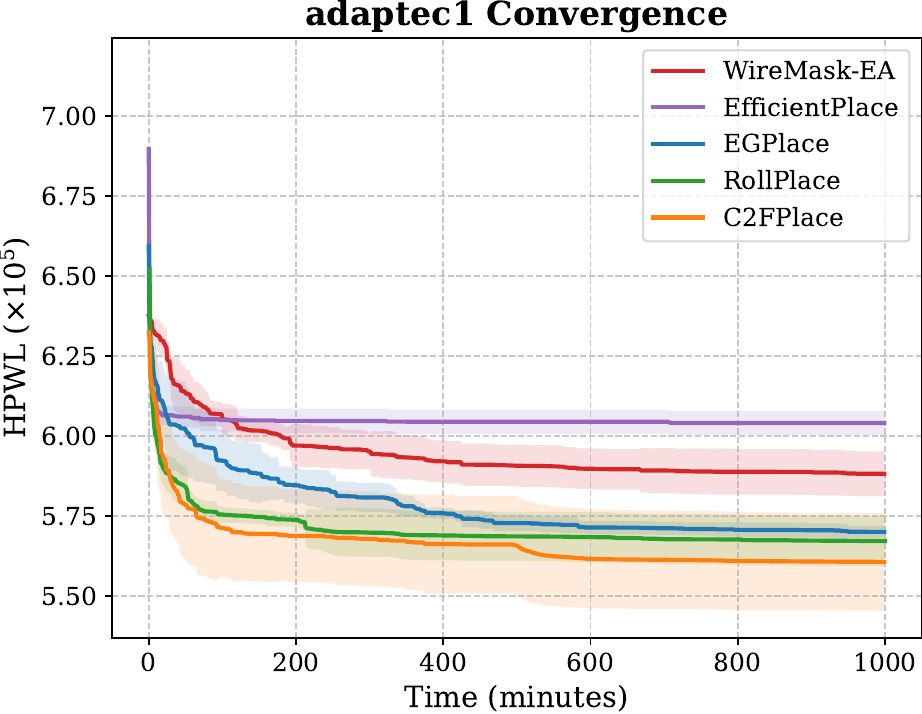} 
    \caption{}
    \label{fig:convergence-adaptec1}
  \end{subfigure}
  \hfill 
  \begin{subfigure}[b]{0.24\textwidth}
    \centering
    \includegraphics[width=\textwidth]{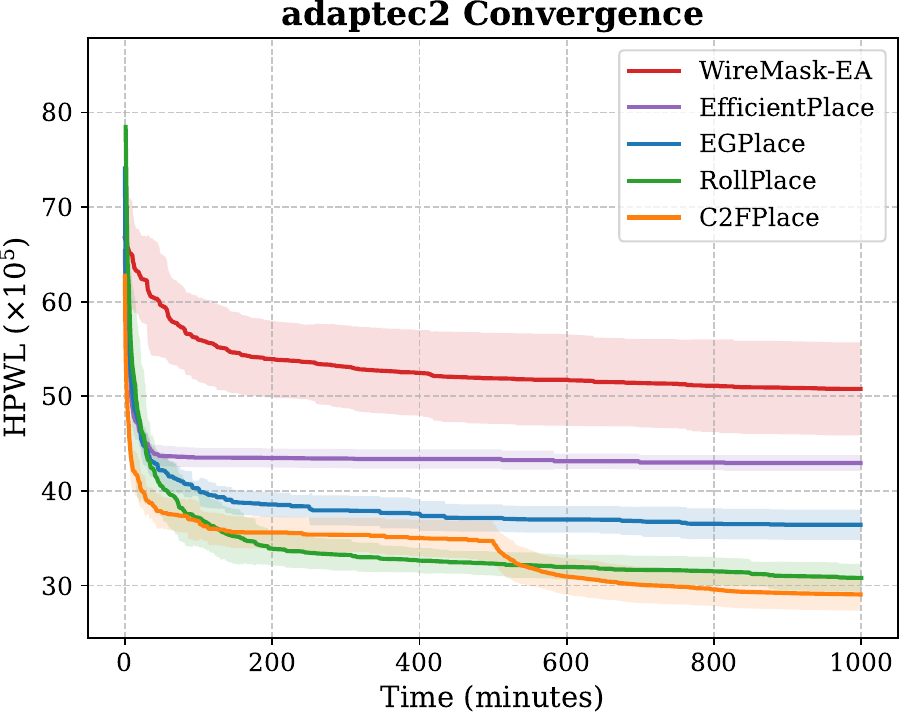} 
    \caption{}
    \label{fig:convergence-adaptec2}
  \end{subfigure}
  \hfill
  \begin{subfigure}[b]{0.24\textwidth}
    \centering
    \includegraphics[width=\textwidth]{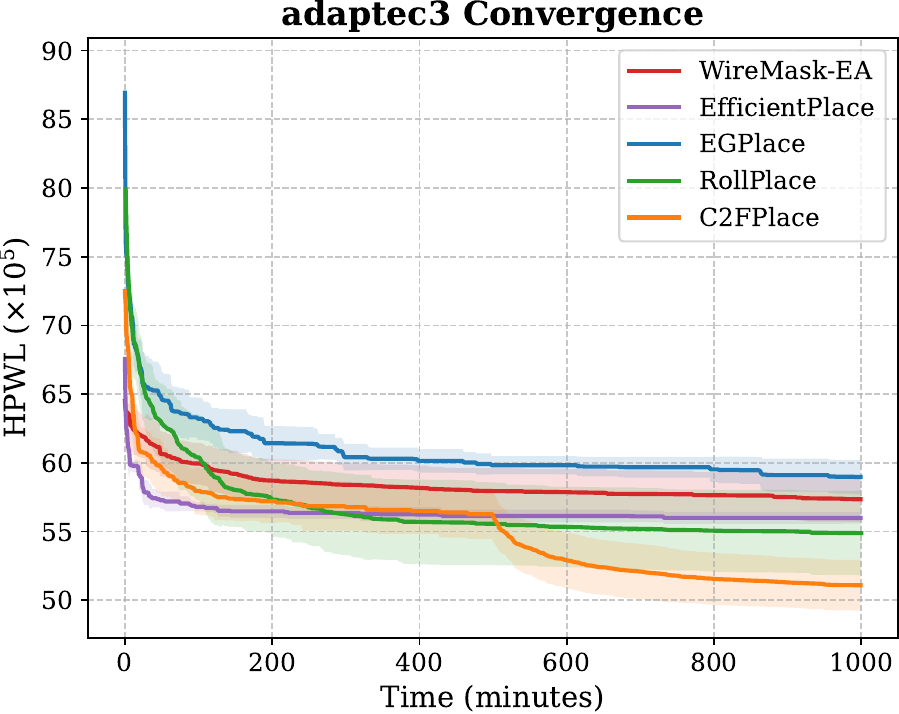} 
    \caption{}
    \label{fig:convergence-adaptec3}
  \end{subfigure}
  \hfill
  \begin{subfigure}[b]{0.24\textwidth}
    \centering
    \includegraphics[width=\textwidth]{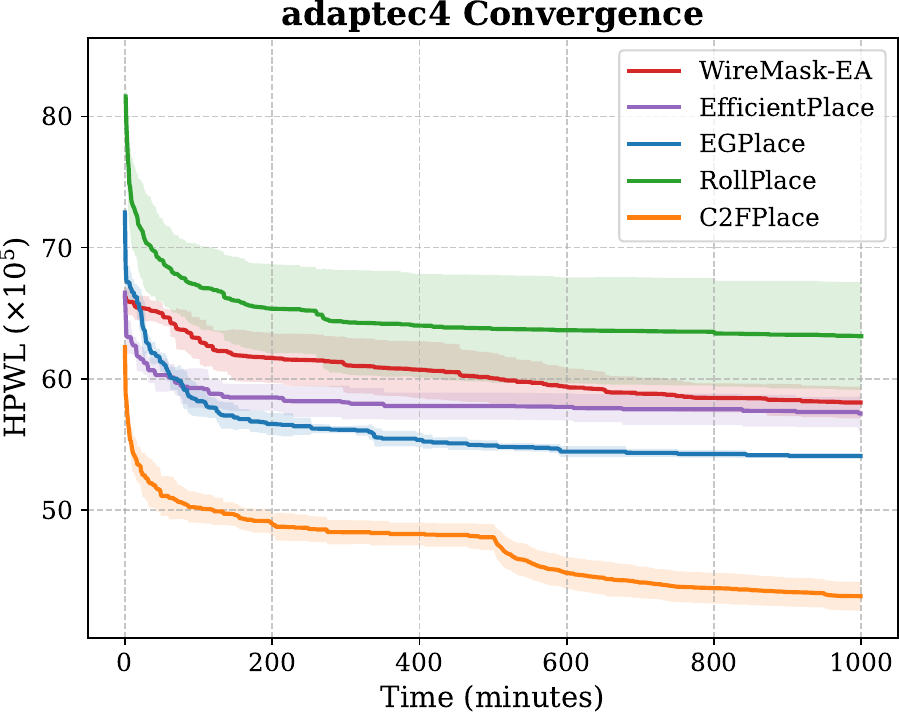}    
    \caption{}
    \label{fig:convergence-adaptec4}
  \end{subfigure}

  \begin{subfigure}[b]{0.24\textwidth}
    \centering
    \includegraphics[width=\textwidth]{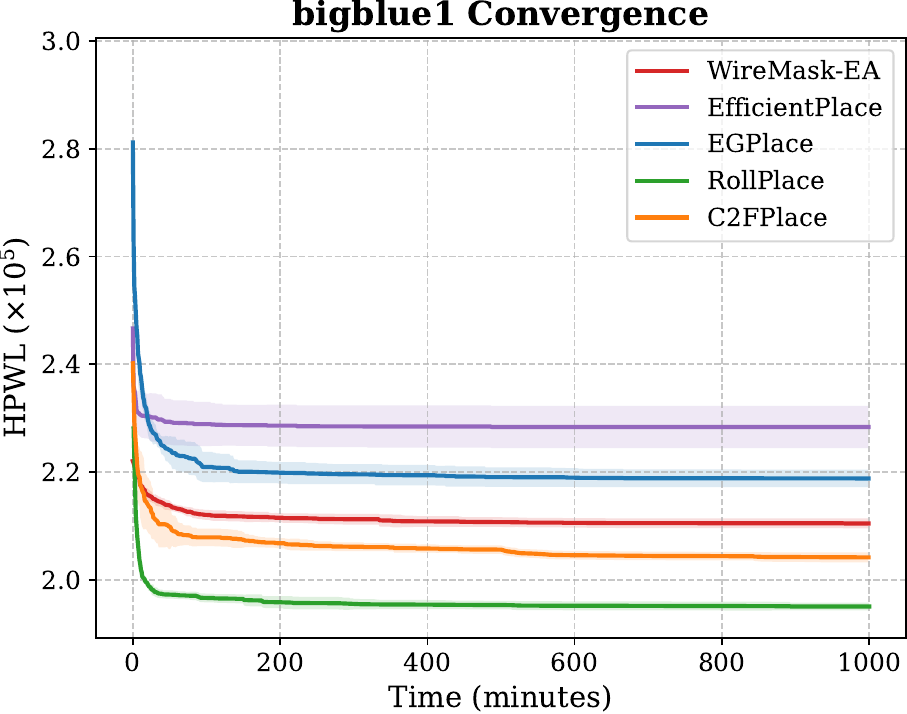} 
    \caption{}
    \label{fig:convergence-bigblue1}
  \end{subfigure}
  \hfill
  \begin{subfigure}[b]{0.24\textwidth}
    \centering
    \includegraphics[width=\textwidth]{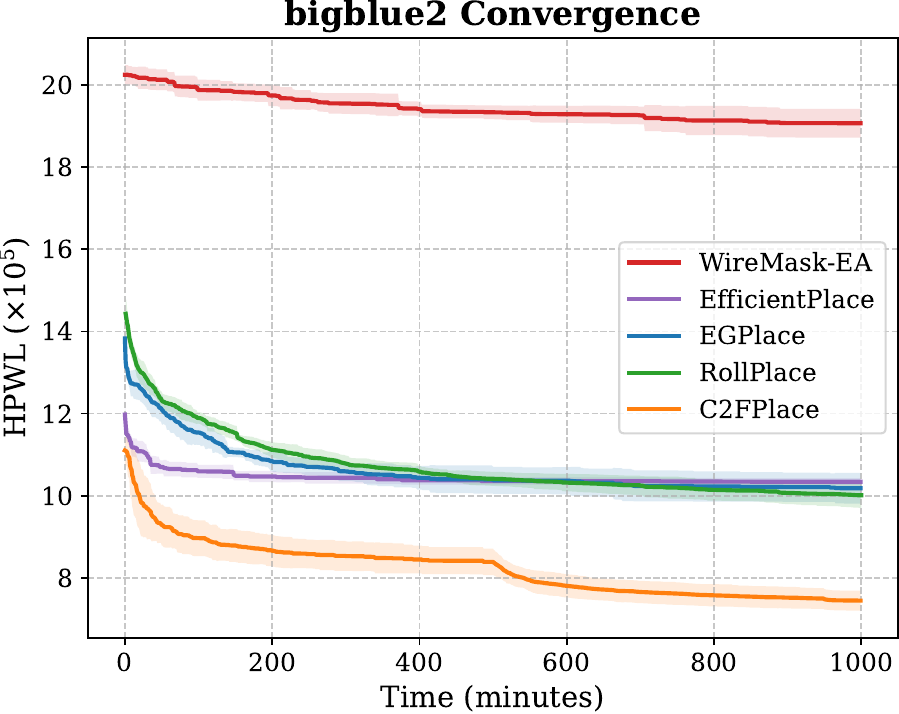} 
    \caption{}
    \label{fig:convergence-bigblue2}
  \end{subfigure}
  \hfill
  \begin{subfigure}[b]{0.24\textwidth}
    \centering
    \includegraphics[width=\textwidth]{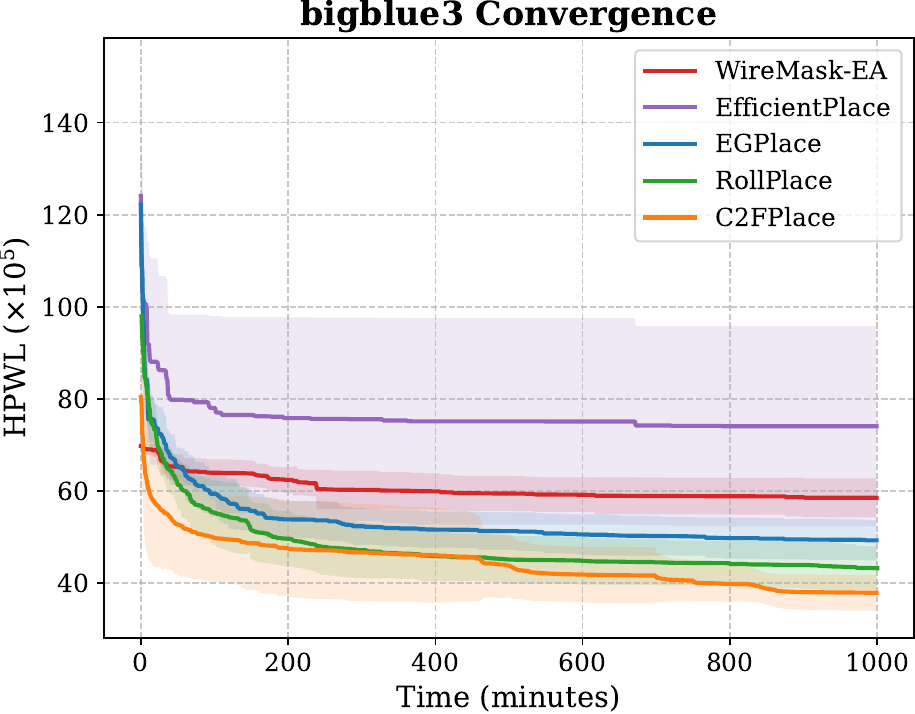} 
    \caption{}
    \label{fig:convergence-bigblue3}
  \end{subfigure}
  \hfill
  \begin{subfigure}[b]{0.24\textwidth}
    \centering
    \includegraphics[width=\textwidth]{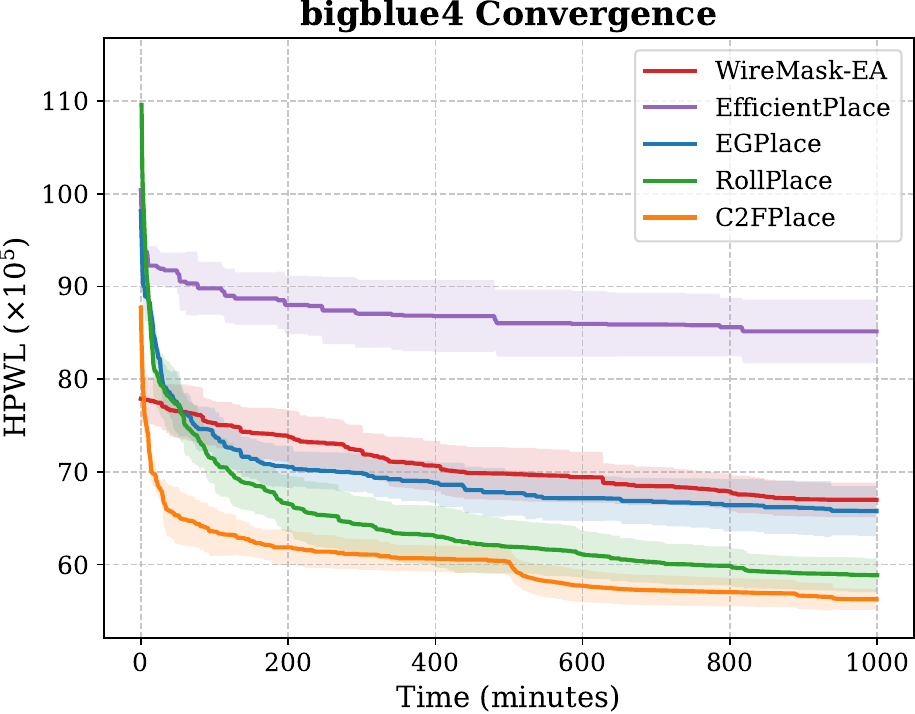}    
    \caption{}
    \label{fig:convergence-bigblue4}
  \end{subfigure}

\caption{Convergence of best-so-far HPWL over runtime for methods rerun on our AMD EPYC 9654 CPU and RTX 6000 ada under a unified 1000-minute budget. Solid lines denote mean values, and shaded regions indicate standard deviation intervals over five independent runs.}
  \label{fig:convergence}
\end{figure}

\subsubsection{Convergence Analysis}
As shown in Figure~\ref{fig:convergence}, all methods are evaluated under the same 1000-minute runtime budget, with all other hyperparameters following their original settings, and C2FPlace allocates this budget as 500 min for Stage~2 Phase~1, 400 min for Stage~2 Phase~2, and 100 min for Stage~3.
C2FPlace rapidly reduces HPWL in the early stage and continues to improve throughout optimization, demonstrating effective global exploration while preserving flexibility for later refinement.

The convergence curves also demonstrate the benefit of staged partial rip-up. By exploring diverse local reconstruction trajectories, C2FPlace avoids premature stagnation and maintains steady HPWL improvements, particularly on large and highly connected circuits. The subsequent low-ratio updates and Stage~3 critical macro tuning further reduce residual wirelength overhead, confirming the effectiveness of the coarse-to-fine optimization strategy.

\begin{table*}[t]
\centering
\caption{Intermediate and PPA metrics on the ICCAD2025 benchmarks with a 1000-minute runtime budget. ``\textit{w/}'' and ``\textit{w/o}'' denote the settings with and without Stage~3 critical macro tuning.}
\label{tab:metrics_comparison}
\resizebox{\textwidth}{!}{ 
\begin{tabular}{ll c c c c c c c}
\toprule
\multicolumn{2}{c}{} & \multicolumn{2}{c}{\textbf{Intermediate Metrics}} & \multicolumn{4}{c}{\textbf{PPA Metrics}} \\
\cmidrule(lr){3-4} \cmidrule(lr){5-8}
\multirow{2}{*}{\textbf{Benchmark}} 
& \multirow{2}{*}{\textbf{Method}} 
& \textbf{Placement} 
& \textbf{Detailed Placement} 
& \multicolumn{3}{c}{\textbf{Timing Performance}} 
& \multirow{2}{*}{\textbf{Power \(\downarrow\)}} \\
\cmidrule(lr){3-3} \cmidrule(lr){4-4} \cmidrule(lr){5-7}
& 
& \textbf{HPWL \(\downarrow\)} 
& \textbf{HPWL \(\downarrow\)} 
& \textbf{WNS \(\uparrow\)} 
& \textbf{TNS \(\uparrow\)} 
& \textbf{NVP \(\downarrow\)} 
& \\
\midrule
\multirow{5}{*}{ac97\_top} 
 & WireMask-EA  & 106125 & 115299 & -350.88 & -262524.84 & \textbf{1889} & \textbf{9.18e-02} \\
 & EGPlace & 84244 & 85090 & -305.20 & -257896.27 & 1897 & 1.16e-01 \\
 & RollPlace & 154668 & 158904 & -484.65 & -272175.41 & \textcolor{brown}{1892} & \textcolor{brown}{1.06e-01} \\
 & C2FPlace \textit{w/o} stage 3& \textcolor{brown}{34778} & \textcolor{brown}{35014} & \textbf{-279.11} & \textbf{-216087.58} & 1893 & 1.34e-01 \\
 & C2FPlace \textit{w/} stage 3& \textbf{33837} & \textbf{34164} & \textcolor{brown}{-292.12} & \textcolor{brown}{-234145.58} & 1893 & 1.35e-01 \\
\midrule
\multirow{5}{*}{aes} 
 & WireMask-EA  & 38660 & 58318 & -177.86 & -48986.96 & 547 & 4.26e-02 \\
 & EGPlace & 32228 & 34538 & -121.73 & -13915.60 & 431 & 3.97e-02 \\
 & RollPlace & 57638 & 68329 & -243.78 & -49069.95 & 564 & 4.31e-02 \\
 & C2FPlace \textit{w/o} stage 3& \textcolor{brown}{28518} & \textcolor{brown}{30546} & \textcolor{brown}{-93.51} & \textcolor{brown}{-10578.63} & \textcolor{brown}{331} & \textcolor{brown}{3.88e-02} \\
 & C2FPlace \textit{w/} stage 3& \textbf{28071} & \textbf{30267} & \textbf{-85.26} & \textbf{-8798.57} & \textbf{329} & \textbf{3.86e-02} \\
\midrule
\multirow{5}{*}{aes\_cipher\_top}
 & WireMask-EA  & 101754 & 119643 & -293.34 & -61544.95 & 596 & 9.77e-02 \\
 & EGPlace & 103438 & 105149 & -168.46 & -32984.75 & 498 & 9.15e-02 \\
 & RollPlace & 323464 & 340744 & -768.79 & -271215.72 & 935 & 1.52e-01 \\
 & C2FPlace \textit{w/o} stage 3& \textcolor{brown}{54767} & \textcolor{brown}{55137} & \textcolor{brown}{-140.75} & \textcolor{brown}{-26973.63} & \textcolor{brown}{394} & \textcolor{brown}{8.15e-02} \\
 & C2FPlace \textit{w/} stage 3& \textbf{53654} & \textbf{54123} & \textbf{-136.97} & \textbf{-25533.20} & \textbf{387} & \textbf{8.12e-02} \\
\midrule
\multirow{5}{*}{des} 
 & WireMask-EA  & 9435 & 13090 & -102.95 & -2972.96 & 44 & 1.73e-02 \\
 & EGPlace & 8224 & 8863 & -94.09 & -2490.54 & \textbf{33} & \textcolor{brown}{1.64e-02} \\
 & RollPlace & 23463 & 26892 & -230.60 & -9451.57 & 95 & 2.10e-02 \\
 & C2FPlace \textit{w/o} stage 3& \textcolor{brown}{7631} & \textcolor{brown}{7890} & \textcolor{brown}{-77.57} & \textcolor{brown}{-2310.32} & \textcolor{brown}{34} & \textbf{1.61e-02} \\
 & C2FPlace \textit{w/} stage 3& \textbf{7476} & \textbf{7781} & \textbf{-77.14} & \textbf{-2273.51} & \textcolor{brown}{34} & \textbf{1.61e-02} \\
\midrule
\multirow{5}{*}{pci\_bridge32}
 & WireMask-EA  & 195019 & 254100 & -782.01 & -1018531.31 & 3250 & 1.72e-01 \\
 & EGPlace & 185046 & 188753 & -612.53 & -555181.25 & 3285 & 1.31e-01 \\
 & RollPlace & 281574 & 305189 & -822.14 & -658746.44 & \textbf{2835} & 1.52e-01 \\
 & C2FPlace \textit{w/o} stage 3& \textcolor{brown}{74777} & \textcolor{brown}{78749} & \textbf{-413.01} & \textbf{-354008.03} & \textcolor{brown}{3151} & \textbf{1.14e-01} \\
 & C2FPlace \textit{w/} stage 3& \textbf{73108} & \textbf{77547} & \textcolor{brown}{-471.76} & \textcolor{brown}{-394838.00} & 3237 & \textcolor{brown}{1.21e-01} \\
 \midrule
\multirow{5}{*}{Average Rank}
 & WireMask-EA  & 3.80 & 4.00 & 4.00 & 4.20 & 3.40 & 3.60 \\
 & EGPlace & 3.20 & 3.00 & 3.00 & 3.00 & 3.40 & \textcolor{brown}{3.00} \\
 & RollPlace & 5.00 & 5.00 & 5.00 & 4.80 & 3.60 & 4.20 \\
 & C2FPlace \textit{w/o} stage 3 & \textcolor{brown}{2.00} & \textcolor{brown}{2.00} & \textcolor{brown}{1.60} & \textcolor{brown}{1.60} & \textcolor{brown}{2.20} & \textbf{2.00} \\
 & C2FPlace \textit{w/} stage 3 & \textbf{1.00} & \textbf{1.00} & \textbf{1.40} & \textbf{1.40} & \textbf{2.00} & \textbf{2.00} \\
\bottomrule
\end{tabular}
}
\end{table*}

\subsection{ICCAD2025 Results}
To assess the scalability and downstream physical design quality of C2FPlace, we further evaluate it on the ICCAD2025 benchmark. We exclude \texttt{ariane} (136 macros) and apply C2FPlace and search-based baselines to standard-cell placement on the remaining five circuits, with the largest design containing over 12K cells. Following the official flow, we use OpenROAD v2.0 with the provided ASAP7 technology library to obtain placement and PPA metrics. Note that Placement HPWL is from the placer, while Detailed Placement HPWL is measured after legalization and detailed placement where the latter is typically larger due to row/site alignment and design-rule constraints.

As shown in Table~\ref{tab:metrics_comparison}, C2FPlace with Stage~3 achieves the lowest placement HPWL on all five circuits, rwith an average
per-circuit reduction of 37.92\% relative to the best baseline for each circuit.  Compared with
C2FPlace without Stage 3, it reduces placement HPWL by 2.11\% on average, and the corresponding
average reduction after detailed placement is 1.62\%.  For downstream PPA, C2FPlace with Stage~3 achieves the best average ranking in WNS (1.40), TNS (1.40), and NVP (2.00), while maintaining competitive power efficiency. These results demonstrate the scalability of C2FPlace and the effectiveness of Stage~3. However, Stage~3 improves average WNS and TNS rankings but degrades timing on \texttt{ac97\_top} and \texttt{pci\_bridge32}, showing that HPWL reduction does not always translate proportionally to downstream timing and power.

\section{Conclusion}
We propose C2FPlace,  a coarse-to-fine macro placement framework that combines tournament
selection, stochastic macro-subset reconstruction, and a two-phase reconstruction-ratio schedule
with critical macro tuning. During the coarse-grained optimization stage, C2FPlace performs evolutionary search through stochastic partial rip-up and re-place, enabling diverse solution exploration while progressively refining promising placement structures. After global optimization, C2FPlace performs fine-grained critical macro tuning to compensate for discretization errors introduced by coarse-grid placement and further improve placement quality. Experiments demonstrate the effectiveness and scalability of C2FPlace. On the ISPD2005 benchmark, C2FPlace achieves the best average HPWL among all compared methods and outperforms the state-of-the-art methods EGPlace and RollPlace by 17.82\% and 17.86\%, respectively. Furthermore, evaluations on the ICCAD2025 benchmark demonstrate that C2FPlace scales to large-scale placement scenarios with thousands to tens of thousands of movable modules and achieves the best overall average ranking in both intermediate placement metrics and downstream PPA evaluations. These results demonstrate that C2FPlace provides an effective and scalable solution for modern macro placement by combining evolutionary search for global exploration with critical macro tuning for local precision optimization.


\bibliography{iclr2027_conference}
\bibliographystyle{iclr2027_conference}

\clearpage
\appendix
\section{Implementation Details}

\subsection{Detailed Optimization Procedures}~\label{app:detail}
This section provides the detailed optimization procedures of the coarse-to-fine refinement process. The complete pseudocode is presented in Algorithms~\ref{alg:stage2} and~\ref{alg:stage3}. The optimization consists of two stages: coarse-grained evolutionary search and fine-grained critical macro tuning.

\begin{algorithm}[htbp]
\caption{Coarse-Grained Evolutionary Search}
\label{alg:stage2}
\begin{algorithmic}[1]

\REQUIRE Initial population $\mathcal{P}$, macro set $\mathcal{M}$, iterations $T$, ratio intervals $R$
\ENSURE Optimized placement $P^\star$

\FOR{$t=1,\ldots,T$}

    \STATE Randomly sample $\mathcal T$ individuals from $\mathcal{P}$

    \STATE Select parent:
    \[
    P_{\mathrm{par}}
    \leftarrow
    \arg\min_{P_i\in \mathcal T}
    \mathrm{HPWL}(P_i)
    \]

    \STATE Sample rip-up ratio $\rho\sim\mathcal{U}(R)$

    \STATE $k\leftarrow\max(1,\lfloor\rho|\mathcal{M}|\rfloor)$

    \STATE Randomly select $k$ macros:
    $\mathcal{S}\leftarrow\mathrm{Sample}(\mathcal{M},k)$

    \STATE Remove $\mathcal{S}$ from $P_{\mathrm{par}}$ to obtain partial layout $\widetilde P$

    \STATE Preserve original relative order:
    $\pi\leftarrow\mathrm{SortByIndex}(\mathcal{S})$
    \FOR{each macro $m\in\pi$}

        \STATE
        $
        c^\star
        \leftarrow
        \arg\min_{c\in\mathcal{L}_c(m|\widetilde P)}
        W_m(c|\widetilde P)
        $

        \STATE Reinsert $m$ into $\widetilde P$

    \ENDFOR

        \STATE Find worst individual:
        \[
        P_{\mathrm{worst}}
        \leftarrow
        \arg\max_{P_i\in\mathcal{P}}
        \mathrm{HPWL}(P_i)
        \]
\IF{$\mathrm{HPWL}(\widetilde P)<\mathrm{HPWL}(P_{\mathrm{worst}})$}
        \STATE Replace $P_{\mathrm{worst}}$ with $\widetilde P$

    \ENDIF

\ENDFOR

\STATE
$P^\star
\leftarrow
\arg\min_{P_i\in\mathcal{P}}
\mathrm{HPWL}(P_i)$

\RETURN $P^\star$

\end{algorithmic}
\end{algorithm}

\begin{algorithm}[htbp]
\caption{Critical Macro Tuning via Structure-Preserving Coarse-Grid Displacement and Fine-Grid Offset Refinement}
\label{alg:stage3}
\begin{algorithmic}[1]

\REQUIRE Initial placement $P$, netlist $\mathcal{N}$,
         iterations $T_3$
\ENSURE Optimized placement $P$

\FOR{$t=1,\ldots,T_3$}
    \STATE $\widetilde{P} \leftarrow P$

    \STATE \textbf{Structure-Preserving Coarse-Grid Displacement}
    \STATE Identify bounding-box critical macros
    $\mathcal{M}_{\mathrm{crit}}$ from $\mathcal{N}$ and $\widetilde{P}$

    \FOR{each macro $m \in \mathcal{M}_{\mathrm{crit}}$}
        \STATE $(\mathbf{c}_m,\boldsymbol{\delta}_m)
        \leftarrow \mathrm{GetPosition}(m,\widetilde{P})$
        \STATE Remove $m$ from $\widetilde{P}$
        \STATE Generate legal coarse locations
        $\mathcal{C}^{\mathrm{coarse}}_m$
        satisfying footprint and structure-preserving constraints
        \STATE $\mathbf{c}_m^\star \leftarrow
        \arg\min_{\mathbf{c}\in\mathcal{C}^{\mathrm{coarse}}_m}
        W_m(\mathbf{c},\boldsymbol{\delta}_m\mid\widetilde{P})$
        \STATE Reinsert $m$ at
        $(\mathbf{c}_m^\star,\boldsymbol{\delta}_m)$
        into $\widetilde{P}$
    \ENDFOR

    \STATE \textbf{Fine-Grid Offset Refinement}
    \STATE Update bounding-box critical macros
    $\mathcal{M}_{\mathrm{crit}}$ from $\mathcal{N}$ and $\widetilde{P}$

    \FOR{each macro $m \in \mathcal{M}_{\mathrm{crit}}$}
        \STATE $(\mathbf{c}_m,\boldsymbol{\delta}_m)
        \leftarrow \mathrm{GetPosition}(m,\widetilde{P})$
        \STATE Remove $m$ from $\widetilde{P}$
        \STATE Generate feasible fine offsets
        $\mathcal{D}^{\mathrm{fine}}_m(\mathbf{c}_m)$
        within the current coarse footprint
        \STATE $\boldsymbol{\delta}_m^\star \leftarrow
        \arg\min_{\boldsymbol{\delta}
        \in\mathcal{D}^{\mathrm{fine}}_m(\mathbf{c}_m)}
        W_m(\mathbf{c}_m,\boldsymbol{\delta}\mid\widetilde{P})$
        \STATE Reinsert $m$ at
        $(\mathbf{c}_m,\boldsymbol{\delta}_m^\star)$
        into $\widetilde{P}$
    \ENDFOR

    \IF{$\mathrm{HPWL}(\widetilde{P}) < \mathrm{HPWL}(P)$}
        \STATE $P \leftarrow \widetilde{P}$
    \ENDIF
\ENDFOR

\RETURN $P$

\end{algorithmic}
\end{algorithm}

\subsection{Benchmark Circuit Statistics}
\label{app:benchmark_statistics}

Table~\ref{tab:iccad25_benchmark_stats} summarizes the circuit statistics of the ISPD2005~\cite{ispd2005} and ICCAD2025~\cite{iccad2025} benchmarks used in our experiments. For the ISPD2005 circuits, C2FPlace optimizes the macro subset, and thus the number of modules to place corresponds to the number of macros. For bigblue2 and bigblue4, we optimize only 1024 macros, following the setting described in the main paper. All macros are placed for the other six ISPD2005 circuits.

For the ICCAD2025 benchmarks, C2FPlace is applied to standard-cell placement, and ``Modules to Place'' therefore denotes the number of standard cells optimized by C2FPlace. The ariane circuit is excluded from the C2FPlace experiments because its placement setting differs from the other ICCAD2025 circuits. The table also reports the numbers of ports, nets, pins, and the corresponding area utilization to provide a detailed overview of the benchmark characteristics.

\begin{table*}[htbp]
  \centering
  \caption{Circuit statistics of the ISPD2005 and ICCAD2025 benchmarks.}
  \label{tab:iccad25_benchmark_stats}
  \resizebox{\textwidth}{!}{ 
  \begin{tabular}{l c c c c c c c}
    \toprule
    \textbf{Circuit} & \textbf{Macros} & \textbf{Standard Cells} & \textbf{Modules to Place} & \textbf{Ports} & \textbf{Nets} & \textbf{Pins} & \textbf{Area Util (\%)} \\
    \toprule
    adaptec1 & 543 & 210904 & 543 & 0 & 221142 & 944053 & 55.62 \\
    adaptec2 & 566 & 254457 & 566 & 0 & 266009 & 1069482 & 74.46 \\
    adaptec3 & 723 & 450927 & 723 & 0 & 466758 & 1875039 & 61.51 \\
    adaptec4 & 1329 & 494716 & 1329 & 0 & 515951 & 1912420 & 48.62 \\
    bigblue1 & 560 & 277604 & 560 & 0 & 284479 & 1144691 & 31.58 \\
    bigblue2 & 23084 & 534782 & 1024 & 0 & 577235 & 2122282 & 32.43 \\
    bigblue3 & 1298 & 1095514 & 1298 & 0 & 1123170 & 3833218 & 66.81 \\
    bigblue4 & 8170 & 2169183 & 1024 & 0 & 2229886 & 8900078 & 35.68 \\
    \midrule
    ac97\_top & 0 & 7750 & 7750 & 132 & 7806 & 28406 & 56.23 \\
    aes & 0 & 4393 & 4393 & 389 & 4653 & 17029 & 57.06 \\
    aes\_cipher\_top & 0 & 11630 & 11630 & 388 & 11890 & 40834 & 64.64 \\
    ariane & 136 & 105594 & 0 & 495 & 108198 & 441645 & 59.89 \\
    des & 0 & 2317 & 2317 & 197 & 2441 & 7549 & 56.09 \\
    pci\_bridge32 & 0 & 12091 & 12091 & 368 & 12254 & 45333 & 59.46 \\
    \bottomrule
  \end{tabular}
  }
\end{table*}

\subsection{Hyperparameter Settings}
\label{app:hyperparameter}

Table~\ref{tab:hyperparameter} summarizes the hyperparameter settings used in C2FPlace. The coarse-grained evolutionary search is performed in two phases with different rip-up ratios. The first phase uses relatively large rip-up ratios to encourage broader exploration, while the second phase gradually reduces the rip-up ratio for more localized refinement. We use a population size of 20 and a tournament size of 5 throughout the evolutionary search.

After the coarse-grained search, we perform 50 iterations of fine-grained critical macro tuning. Unless otherwise specified, all experiments use a coarse grid size of 512 and five random seeds $\{0,1,2,3,4\}$. The experiments are conducted on an AMD EPYC 9654 CPU.

\begin{table}[t]
\centering
\caption{Hyperparameter settings of C2FPlace.}
\label{tab:hyperparameter}
\begin{tabular}{lc}
\toprule
\textbf{Hyperparameter} & \textbf{Value} \\
\midrule
Population size & 20 \\
Tournament size & 5 \\
Stage 2 Phase 1 iterations & 5,000 \\
Stage 2 Phase 1 rip-up ratio & $[0.4,0.7]$ \\
Stage 2 Phase 2 iterations & 5,000 \\
Stage 2 Phase 2 rip-up ratio & $[0.1,0.4]$ \\
Critical macro tuning iterations & 50 \\
Coarse grid size & 512 \\
Random seeds & $\{0,1,2,3,4\}$ \\
Hardware & AMD EPYC 9654 CPU \\
\bottomrule
\end{tabular}
\end{table}

\subsection{PPA Evaluation Flow}
\label{app:ppa_evaluation}

We use OpenROAD v2.0 to evaluate the downstream PPA performance of the placements generated by C2FPlace. For each placement, the corresponding DEF and SDC files are loaded together with the ASAP7 technology and standard-cell libraries. We then perform detailed placement and estimate the parasitics based on the resulting placement. Finally, we report the total negative slack (TNS), worst negative slack (WNS), power, and the number of endpoints with setup violations (NVP). The complete Tcl script used for this evaluation is shown in Fig.~\ref{fig:openroad_ppa}, ensuring a consistent evaluation flow across all tested circuits.

\begin{figure}[htbp]
\centering
\begin{tcolorbox}[
    colback=gray!5,
    colframe=gray!40,
    boxrule=0.5pt,
    arc=2pt,
    left=5pt,
    right=5pt,
    top=5pt,
    bottom=5pt
]
\begin{lstlisting}[
language=tcl,
basicstyle=\ttfamily\scriptsize,
breaklines=true,
columns=fullflexible,
showstringspaces=false,
frame=none
]
set def_file $::env(DEF_FILE)
set sdc_file $::env(SDC_FILE)

foreach libFile [glob "../ASAP7/LIB/*nldm*.lib"] {
    read_liberty $libFile
}

read_lef ../ASAP7/techlef/asap7_tech_1x_201209.lef

foreach lef [glob "../ASAP7/LEF/*.lef"] {
    read_lef $lef
}

read_def $def_file
read_sdc $sdc_file

source ../ASAP7/setRC.tcl

detailed_placement
estimate_parasitics -placement

report_tns
report_wns
report_power

set nvp_setup [sta::endpoint_violation_count max]
puts "NVP (setup violation count): $nvp_setup"
\end{lstlisting}
\end{tcolorbox}

\caption{OpenROAD v2.0 Tcl script for PPA evaluation.}
\label{fig:openroad_ppa}
\end{figure}

\section{Additional Results}

To evaluate the sensitivity of C2FPlace to key design choices, we conduct an ablation study on eight ISPD2005 benchmark~\cite{ispd2005} circuits, focusing on coarse-grid granularity and rip-up ratio scheduling. We compare the proposed staged random-ratio schedule with ten fixed-ratio settings ranging from 0.1 to 1.0. Each heatmap cell in Fig.~\ref{fig:ratio-coarse-grid-heatmaps} reports the mean final-best HPWL over five random seeds.

To aggregate results across circuits, we use normalized HPWL instead of directly averaging raw HPWL values, since circuit scales vary significantly across the benchmark suite. For circuit $c$, coarse-grid size $g$, and ratio setting $s$, let $H_{c,g,s}$ denote the five-seed mean HPWL. The normalized metric is defined as:
\begin{equation}
\bar{R}_{g,s}
=
\frac{1}{|\mathcal{C}|}
\sum_{c\in\mathcal{C}}
\frac{H_{c,g,s}}
{\min_{s'\in\mathcal{S}}H_{c,g,s'}} ,
\end{equation}
where $\mathcal{C}$ denotes the eight circuits and $\mathcal{S}$ contains all ratio settings, including fixed ratios and the staged random schedule. A lower $\bar{R}_{g,s}$ indicates better overall performance, while $\bar{R}_{g,s}-1$ represents the average relative degradation from the best setting for each circuit.

\subsection{Impact of Coarse-grid Granularity}
\begin{figure*}[t]
  \centering
  \captionsetup[subfigure]{skip=1pt}
  \begin{subfigure}[t]{0.49\textwidth}
    \centering
    \includegraphics[width=\linewidth]{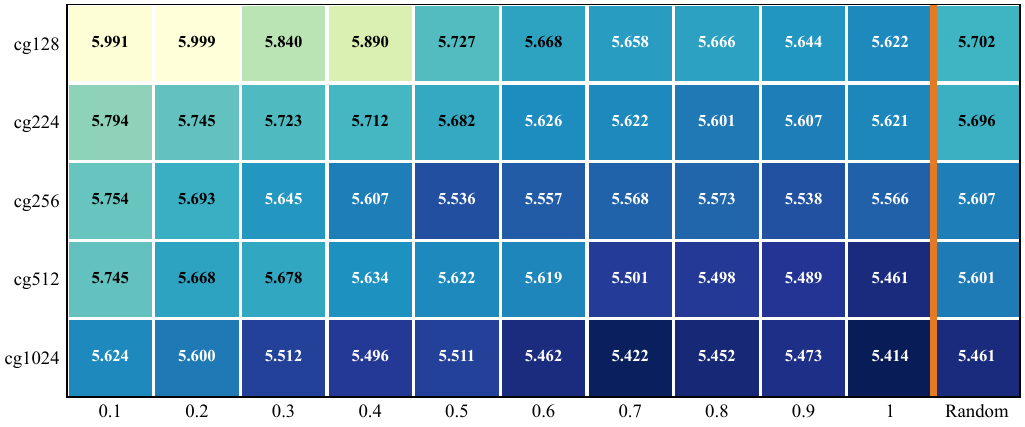}
    \caption{adaptec1 ($\times 10^{5}$)}
  \end{subfigure}\hfill
  \begin{subfigure}[t]{0.49\textwidth}
    \centering
    \includegraphics[width=\linewidth]{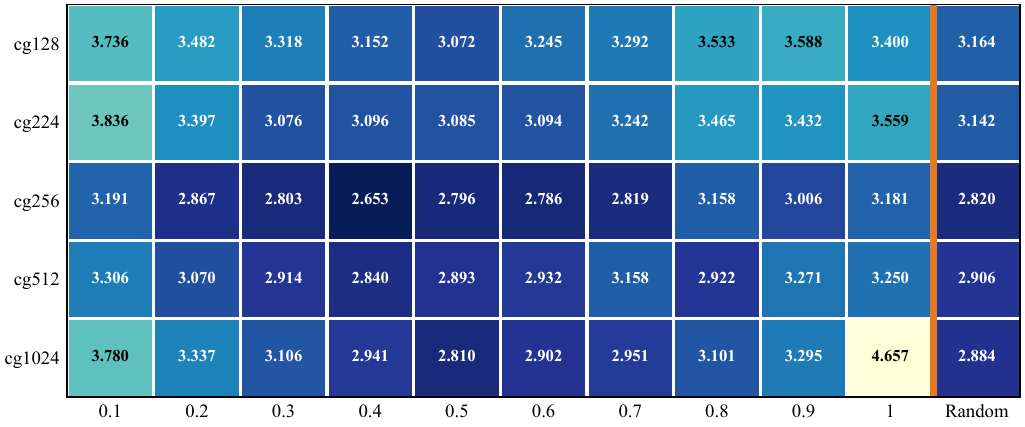}
    \caption{adaptec2 ($\times 10^{6}$)}
  \end{subfigure}

  \begin{subfigure}[t]{0.49\textwidth}
    \centering
    \includegraphics[width=\linewidth]{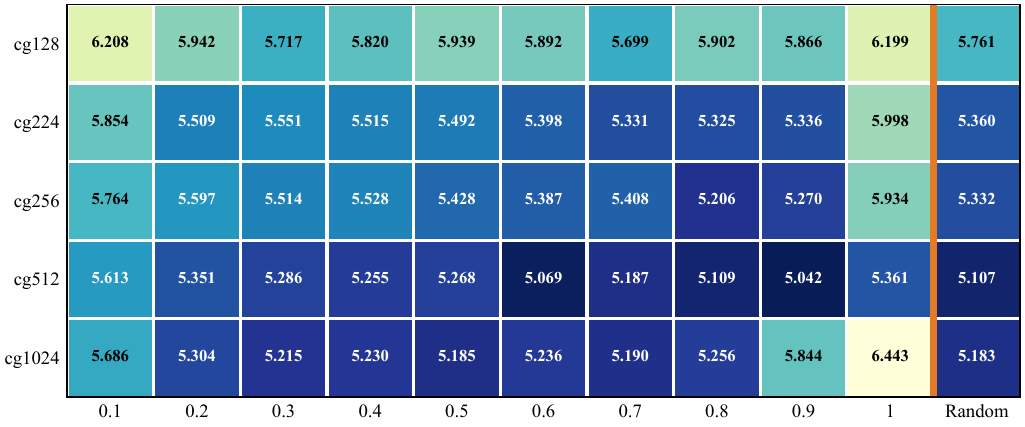}
    \caption{adaptec3 ($\times 10^{6}$)}
  \end{subfigure}\hfill
  \begin{subfigure}[t]{0.49\textwidth}
    \centering
    \includegraphics[width=\linewidth]{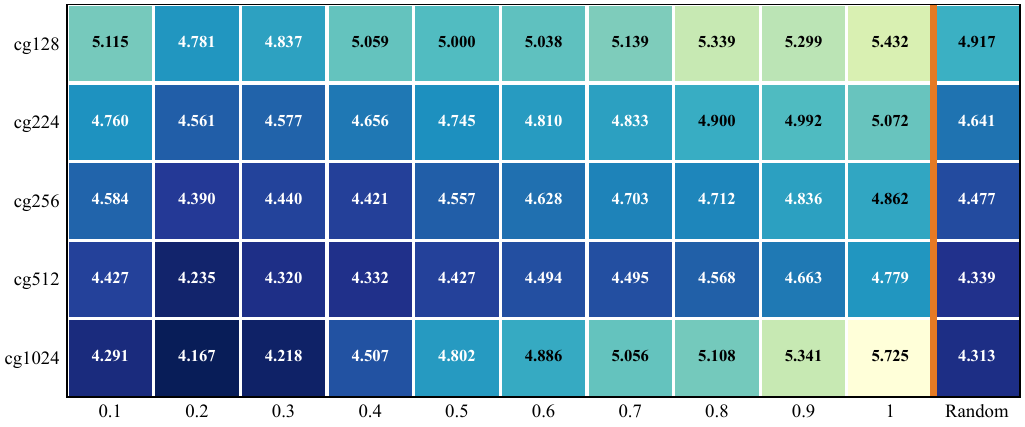}
    \caption{adaptec4 ($\times 10^{6}$)}
  \end{subfigure}

  \begin{subfigure}[t]{0.49\textwidth}
    \centering
    \includegraphics[width=\linewidth]{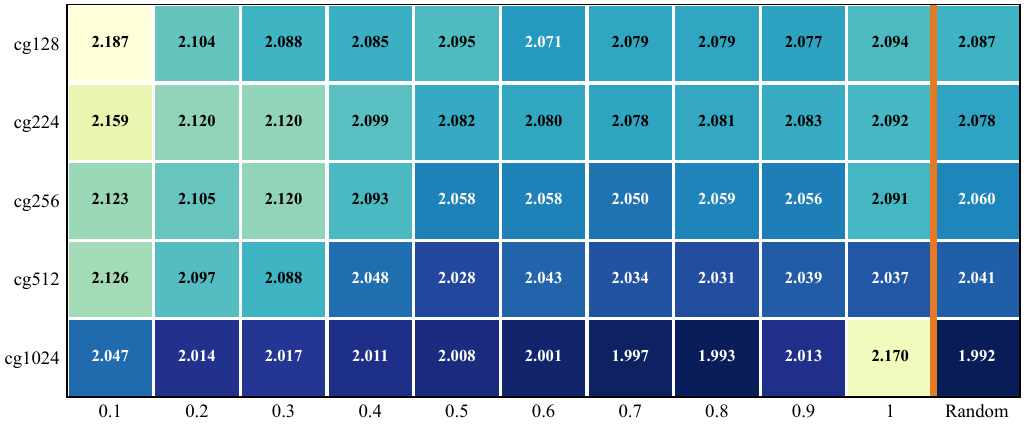}
    \caption{bigblue1 ($\times 10^{5}$)}
  \end{subfigure}\hfill
  \begin{subfigure}[t]{0.49\textwidth}
    \centering
    \includegraphics[width=\linewidth]{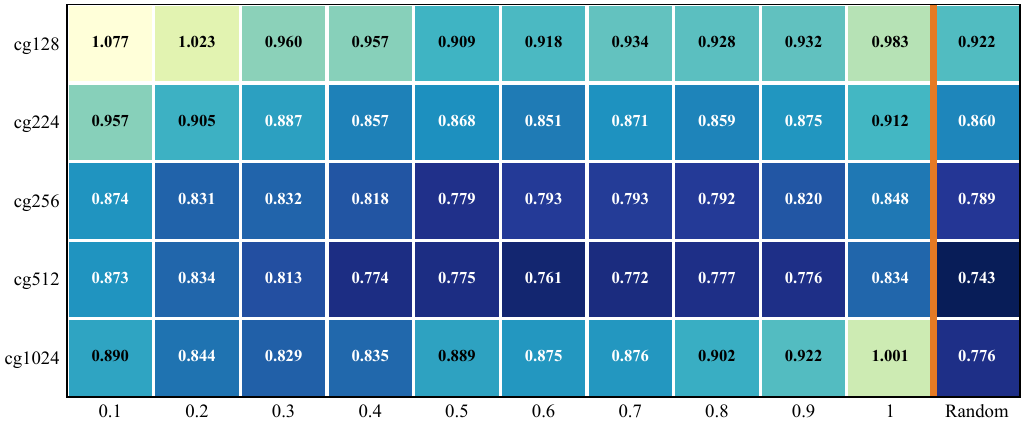}
    \caption{bigblue2 ($\times 10^{6}$)}
  \end{subfigure}

  \begin{subfigure}[t]{0.49\textwidth}
    \centering
    \includegraphics[width=\linewidth]{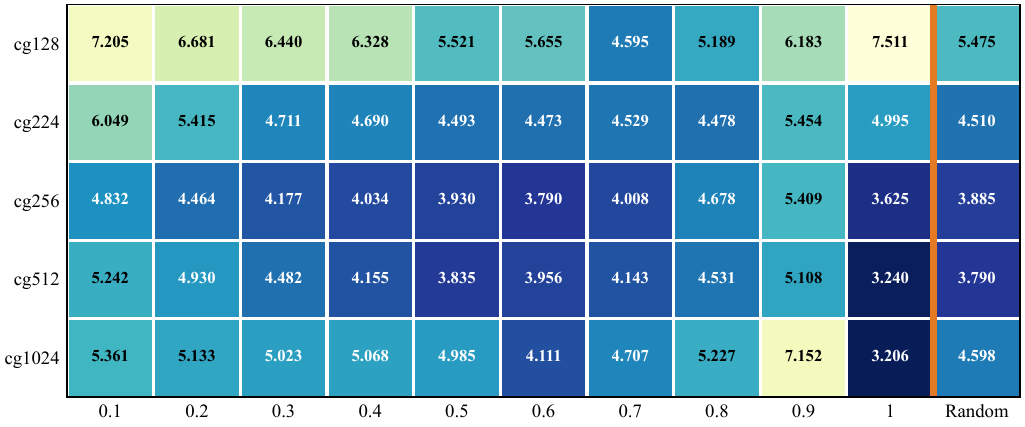}
    \caption{bigblue3 ($\times 10^{6}$)}
  \end{subfigure}\hfill
  \begin{subfigure}[t]{0.49\textwidth}
    \centering
    \includegraphics[width=\linewidth]{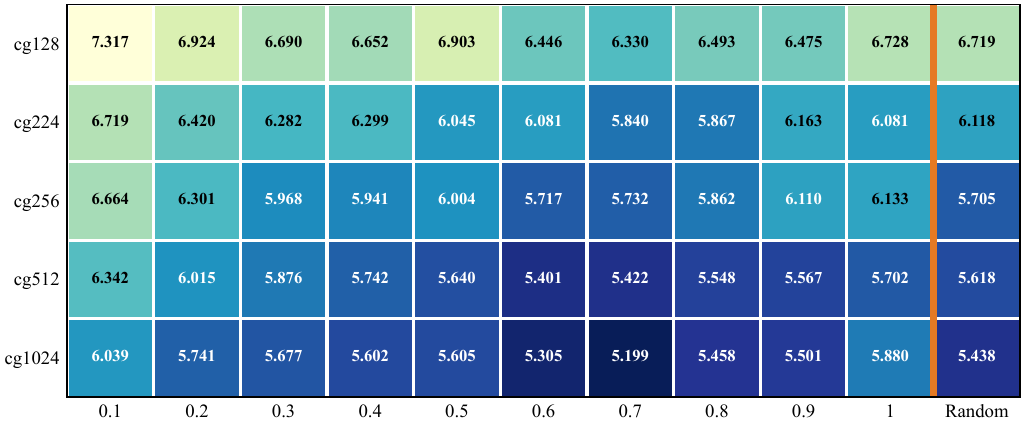}
    \caption{bigblue4 ($\times 10^{6}$)}
  \end{subfigure}

  \caption{Mean final-best HPWL over five seeds for fixed and random ratio settings. Columns in each panel correspond to fixed-ratio settings, followed by the random-ratio strategy after the orange separator; rows correspond to coarse-grid sizes.}
  \label{fig:ratio-coarse-grid-heatmaps}
\end{figure*}

We first investigate the influence of coarse-grid granularity on placement quality. As shown in Fig.~\ref{fig:ratio-coarse-grid-heatmaps}, increasing the grid size generally improves HPWL by providing finer placement resolution. However, the improvement is not monotonic and varies across circuits and rip-up ratios. The numerical scales are reported separately in the subcaptions, where \texttt{adaptec1} and \texttt{bigblue1} use $\times 10^5$ while the remaining circuits use $\times 10^6$.

The optimal granularity is circuit dependent. For example, \texttt{adaptec2} achieves its best result at \texttt{cg256}, while \texttt{adaptec3} and \texttt{bigblue2} obtain the lowest HPWL at \texttt{cg512}; the remaining circuits prefer \texttt{cg1024}. These results indicate that finer grids do not always guarantee better placement quality, as the effectiveness of granularity depends on the interaction between resolution and perturbation strength. Therefore, coarse-grid granularity should be considered as an important component of the coarse-to-fine optimization process rather than a fixed parameter shared across all circuits.

\subsection{Effectiveness of Random Rip-up Ratio}

We evaluate the staged random-ratio schedule using the normalized mean HPWL. The random schedule achieves the best performance at \texttt{cg256}, \texttt{cg512}, and \texttt{cg1024}, with normalized HPWL values of 1.026, 1.038, and 1.069, respectively. Its advantage becomes more evident on finer grids: at \texttt{cg512} and \texttt{cg1024}, it outperforms the best-performing fixed ratio 0.6 (1.038 vs. 1.048 and 1.069 vs. 1.082). For coarse grids, fixed ratios remain slightly competitive, with fixed ratio 0.7 achieving the best result on \texttt{cg128} and fixed ratio 0.6 on \texttt{cg224}.

Averaged over all coarse-grid sizes, the staged random schedule achieves the lowest normalized HPWL of 1.039, followed by fixed ratio 0.6 with 1.044. These results demonstrate that the staged schedule provides a robust performance across different grid granularities without requiring circuit-specific ratio tuning. While individual fixed ratios may still achieve better results on specific cases, the random schedule offers a strong overall trade-off under a unified optimization strategy.

\subsection{Analysis of Ratio Scheduling Strategy}

The fixed-ratio results show that no single ratio consistently achieves the best HPWL across all benchmark circuits. The optimal fixed ratio varies significantly, ranging from 0.2 on \texttt{adaptec4} to 1.0 on \texttt{adaptec1} and \texttt{bigblue3}, indicating that the preferred perturbation strength depends on circuit characteristics and grid granularity. It should be noted that for all fixed‑ratio baselines, the rip‑up ratio in Stage~2 Phase~2 is set to half of that used in Stage~2 Phase~1.

C2FPlace addresses this sensitivity with a staged random-ratio schedule. The global-exploration stage samples rip-up ratios from $[0.4,0.7]$, while the local-refinement stage uses $[0.1,0.4]$. This design provides stronger perturbations during exploration and more conservative updates during refinement, while avoiding dependence on a manually selected fixed ratio.

Overall, the normalized results demonstrate that the staged schedule achieves the best average performance across different coarse-grid sizes, particularly on \texttt{cg256}, \texttt{cg512}, and \texttt{cg1024}. Although fixed ratios remain competitive on \texttt{cg128} and \texttt{cg224}, the proposed schedule provides a robust default across diverse benchmarks without circuit-specific tuning.

\subsection{Visualization Results}\label{app:visual}

\begin{figure*}[tbp]
  \centering
  \begin{subfigure}[b]{0.135\textwidth}
    \centering
    \includegraphics[width=\linewidth]{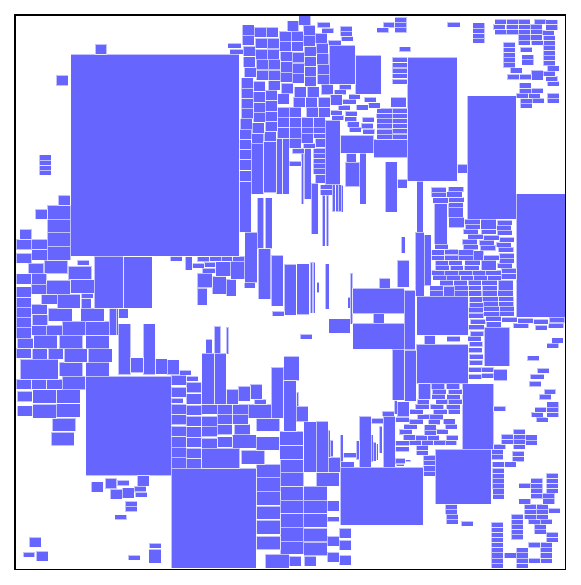}
    \caption{DMPlace \\ \(7.239\times 10^6\)}
  \end{subfigure}\hfill
  \begin{subfigure}[b]{0.135\textwidth}
    \centering
    \includegraphics[width=\linewidth]{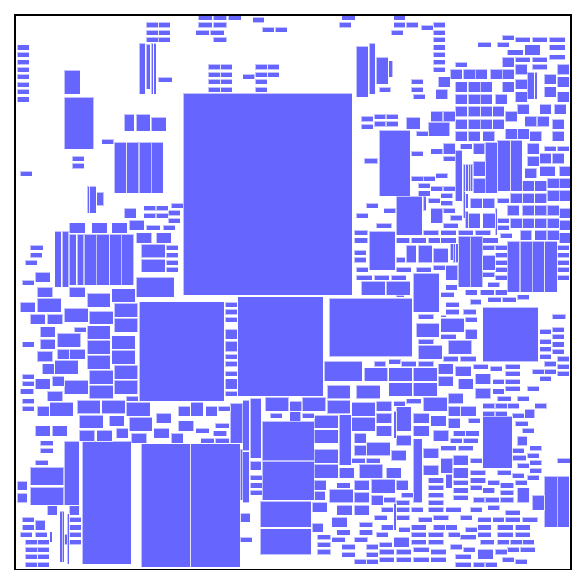}
    \caption{MaskPlace \\ \(8.151 \times 10^6\)}
  \end{subfigure}\hfill
  \begin{subfigure}[b]{0.135\textwidth}
    \centering
    \includegraphics[width=\linewidth]{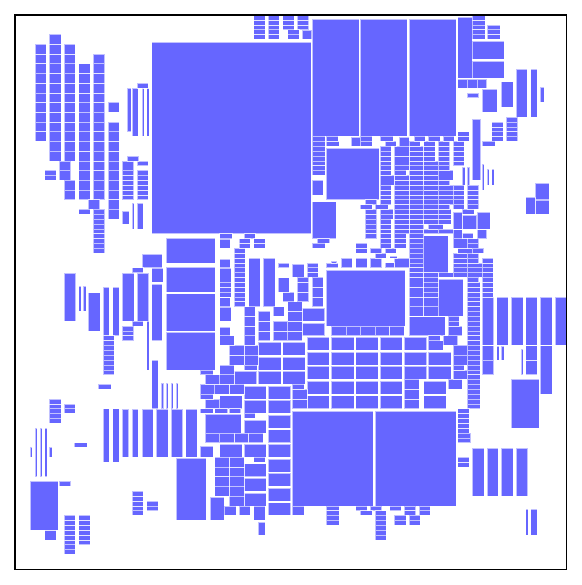}
    \caption{WireM-EA \\ \(5.864 \times 10^6\)}
  \end{subfigure}\hfill
  \begin{subfigure}[b]{0.135\textwidth}
    \centering
    \includegraphics[width=\linewidth]{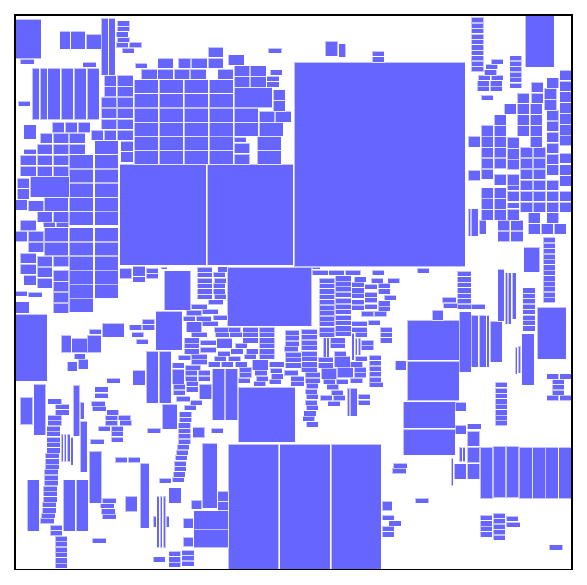}
    \caption{EffiPlace\\ \(5.551 \times 10^6\)}
  \end{subfigure}\hfill
  \begin{subfigure}[b]{0.135\textwidth}
    \centering
    \includegraphics[width=\linewidth]{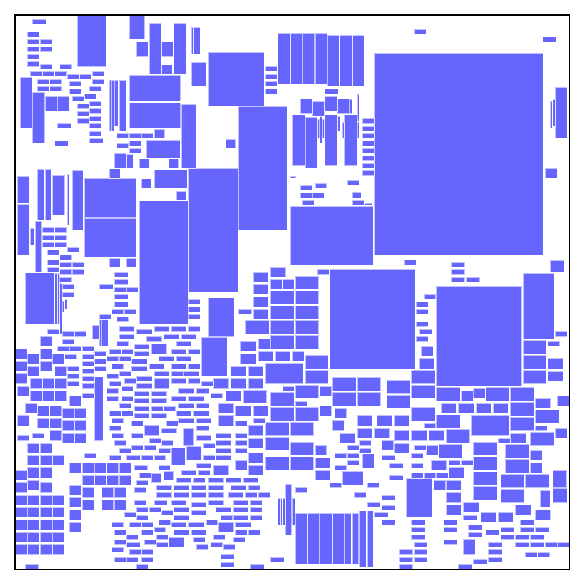}
    \caption{EGPlace \\ \(5.761 \times 10^6\)}
  \end{subfigure}\hfill
  \begin{subfigure}[b]{0.135\textwidth}
    \centering
    \includegraphics[width=\linewidth]{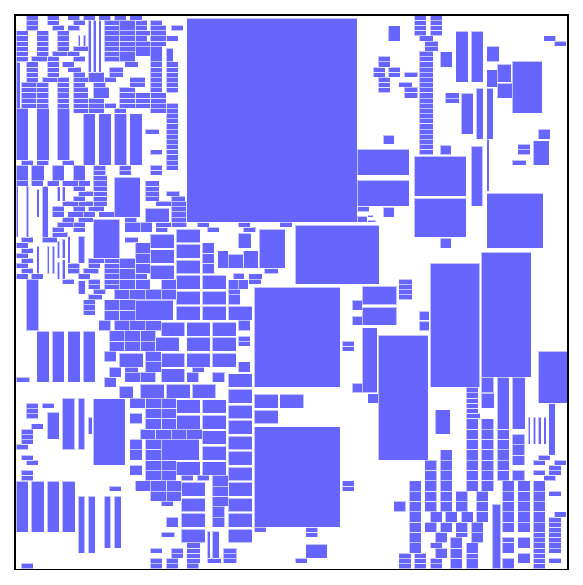}
    \caption{RollPlace \\ \textbf{\(5.251 \times 10^6\)}}
  \end{subfigure}\hfill
  \begin{subfigure}[b]{0.135\textwidth}
    \centering
    \includegraphics[width=\linewidth]{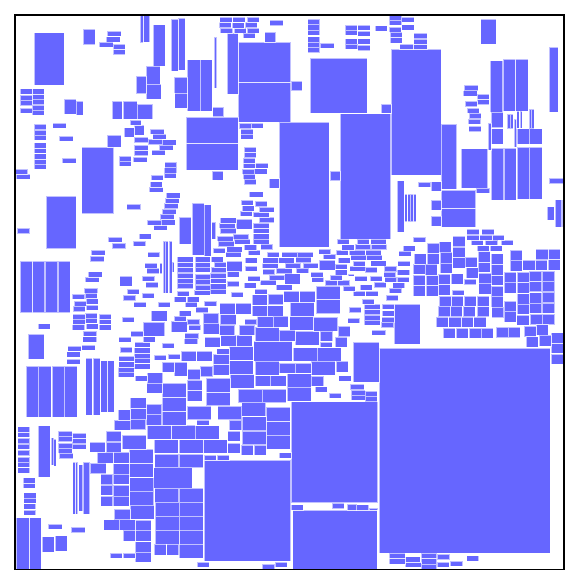}
    \caption{\textbf{C2FPlace} \\ \textbf{\(4.791 \times 10^6\)}}
  \end{subfigure}
  \caption{Macro placement visualization for \texttt{adaptec3} circuit comparing C2FPlace with five recent methods. Abbreviations: DMPlace $=$ DREAMPlace, WireM-EA $=$ WireMask-EA, EffiPlace $=$ EfficientPlace. Blue rectangles denote macros. The placement quality is evaluated through HPWL value, where lower HPWL values indicate superior performance.}
  \label{fig:macro_placement}
\end{figure*}

Figure~\ref{fig:macro_placement} presents a visual comparison of macro placements produced by different methods on \texttt{adaptec3}.

\begin{figure}[t]
\centering

\begin{subfigure}[b]{0.19\textwidth}
    \centering
    \includegraphics[width=\linewidth]{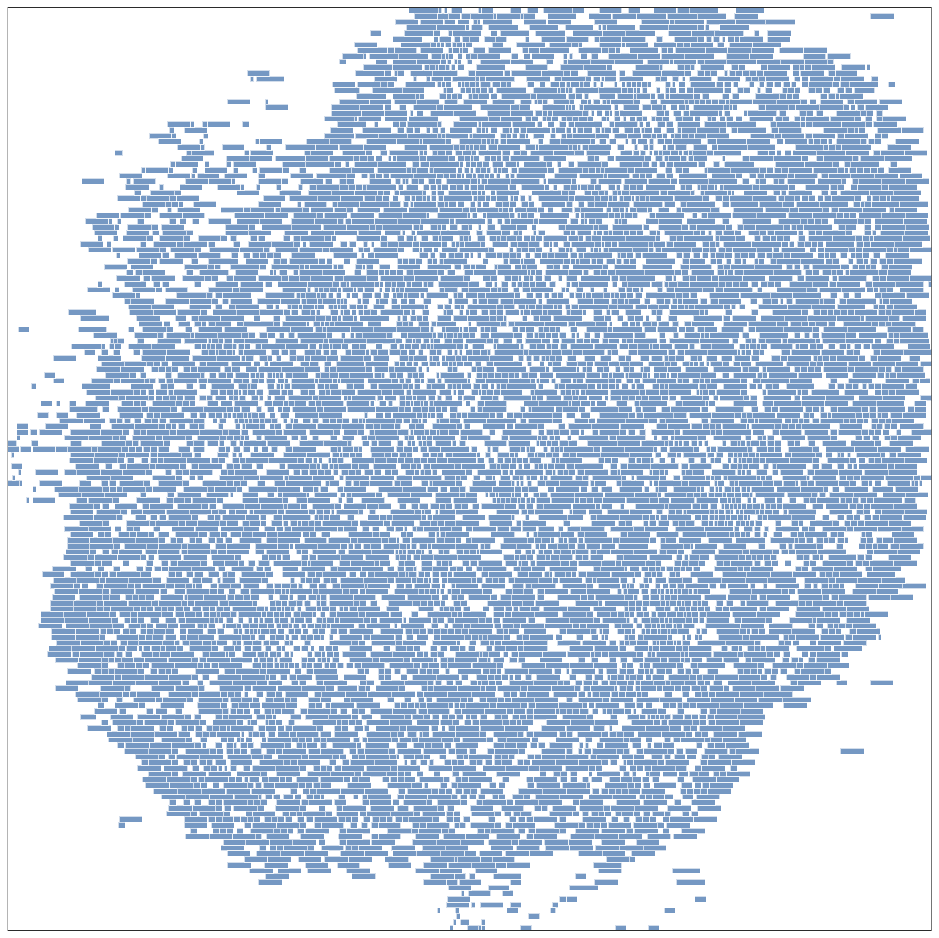}
    \caption{ac97\_top}
    \label{fig:ac97_top_stage2}
\end{subfigure}
\hfill
\begin{subfigure}[b]{0.19\textwidth}
    \centering
    \includegraphics[width=\linewidth]{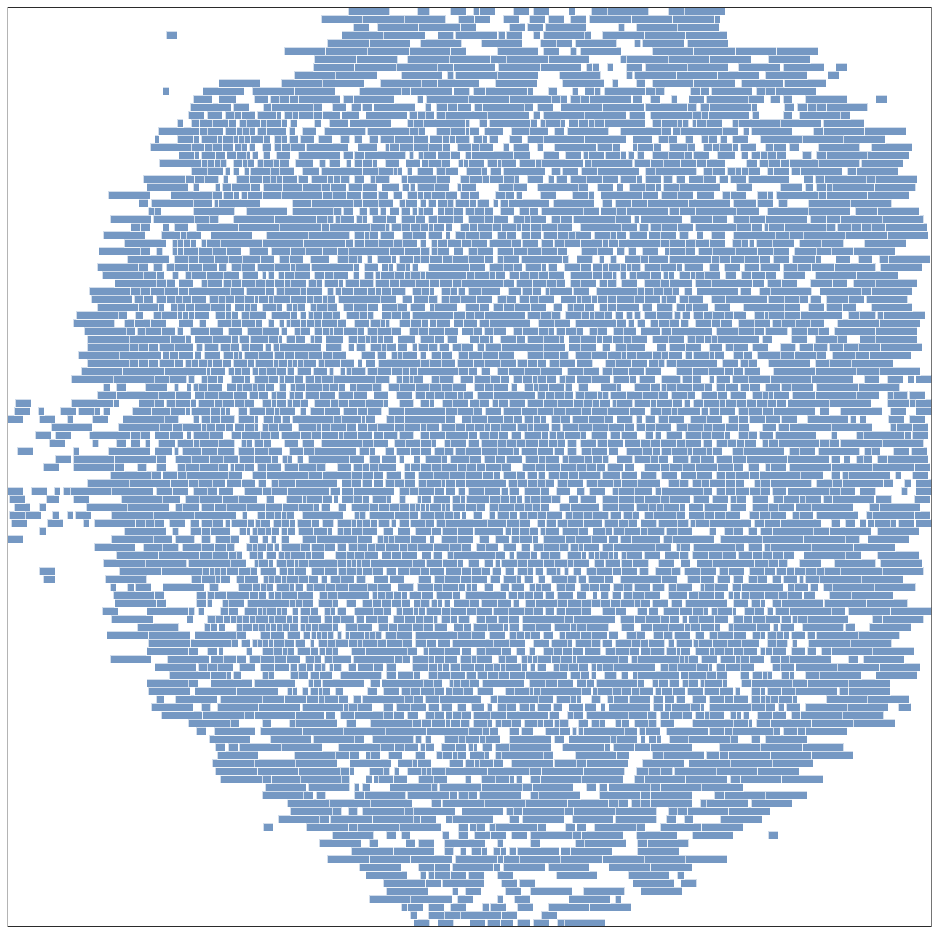}
    \caption{aes}
    \label{fig:aes_stage2}
\end{subfigure}
\hfill
\begin{subfigure}[b]{0.19\textwidth}
    \centering
    \includegraphics[width=\linewidth]{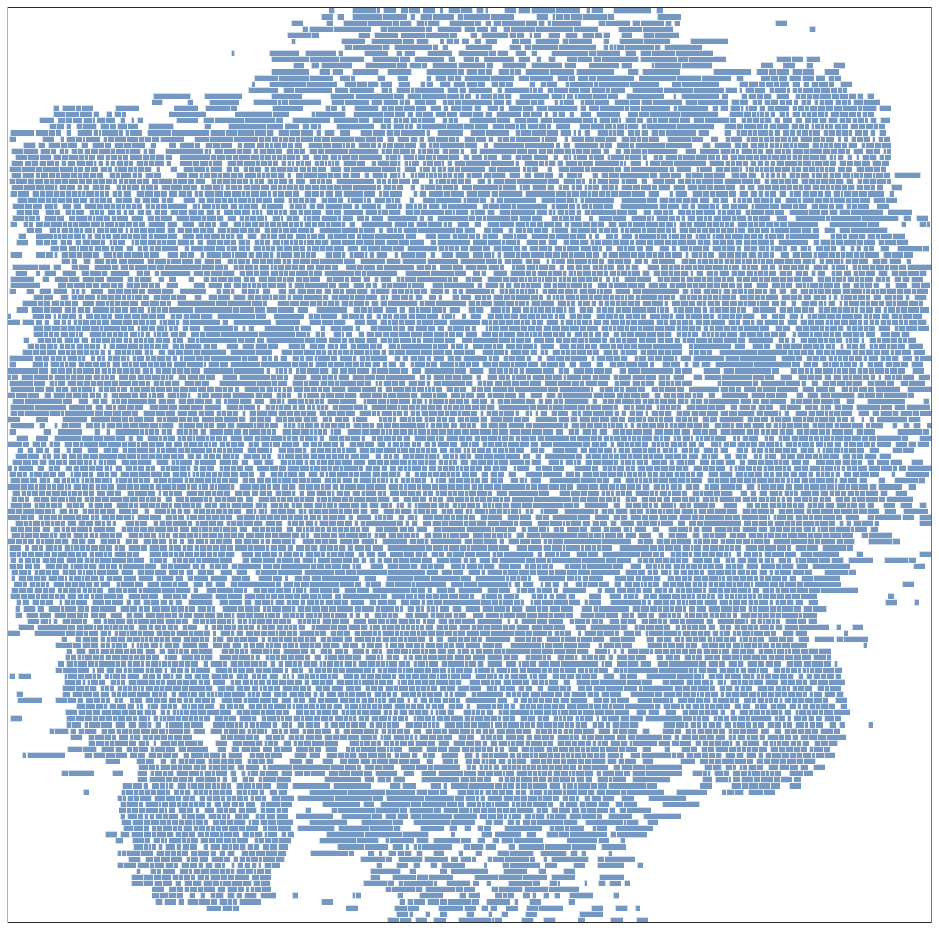}
    \caption{aes\_cipher\_top}
    \label{fig:aes_cipher_top_stage2}
\end{subfigure}
\hfill
\begin{subfigure}[b]{0.19\textwidth}
    \centering
    \includegraphics[width=\linewidth]{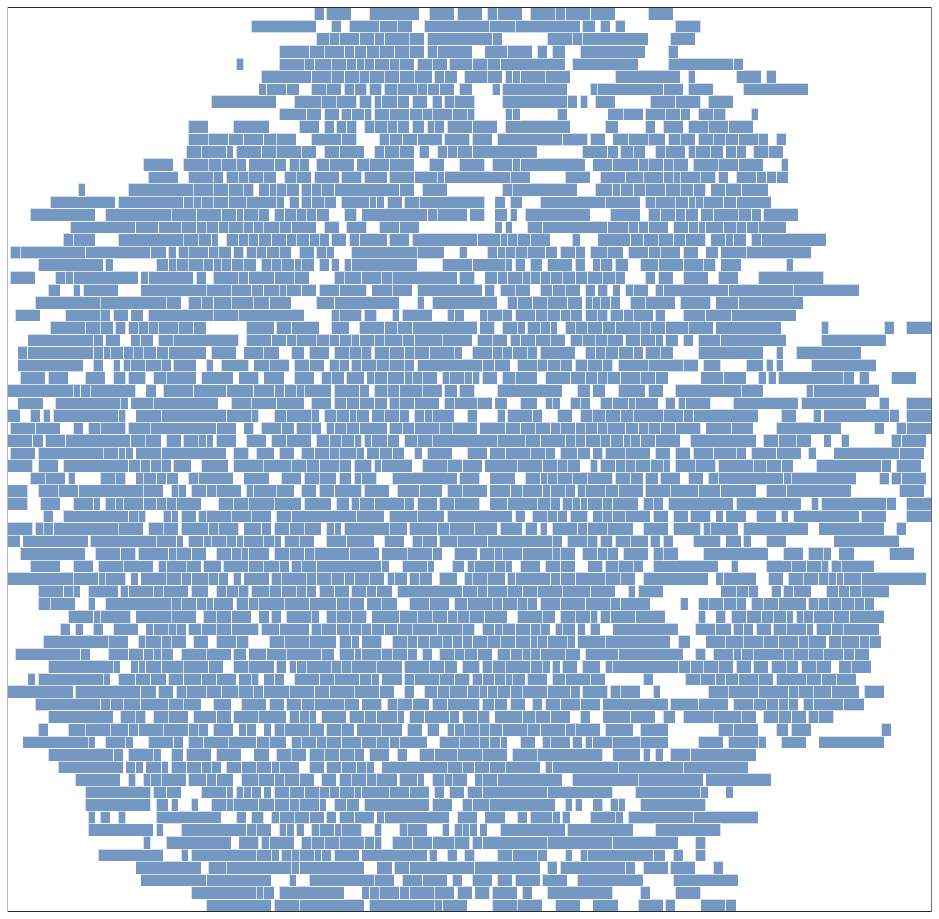}
    \caption{des}
    \label{fig:des_stage2}
\end{subfigure}
\hfill
\begin{subfigure}[b]{0.19\textwidth}
    \centering
    \includegraphics[width=\linewidth]{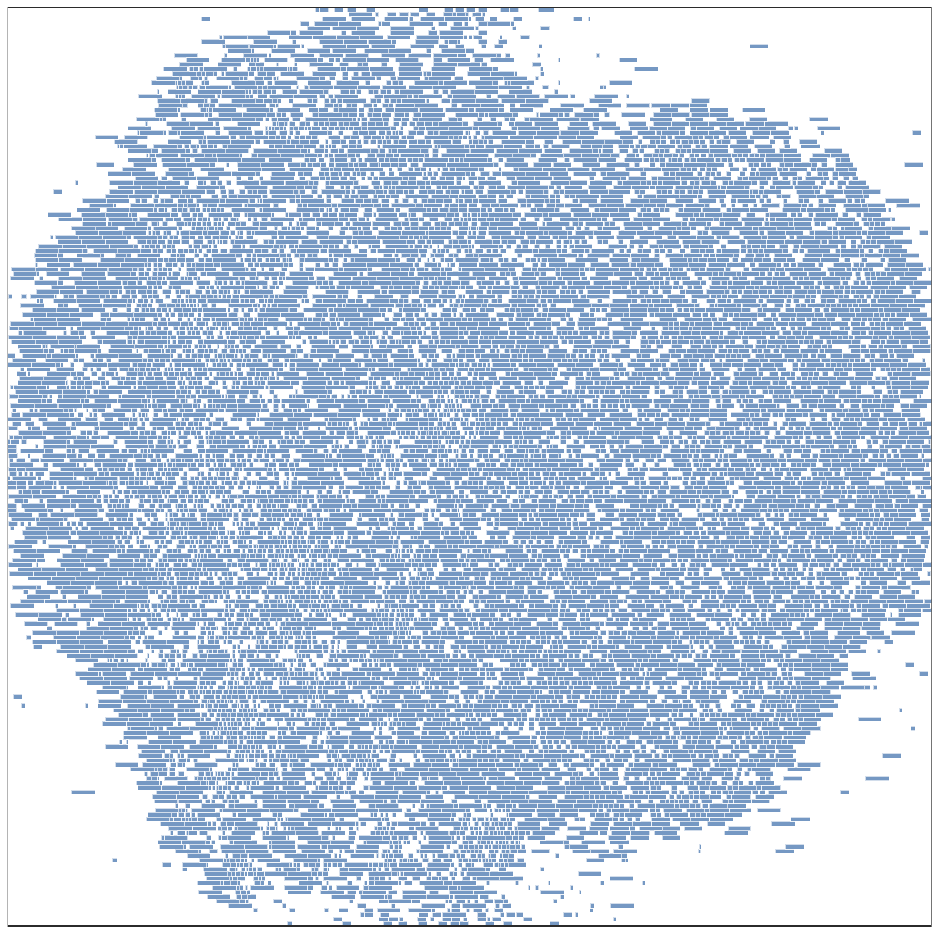}
    \caption{pci\_bridge32}
    \label{fig:pci_bridge32_stage2}
\end{subfigure}

\begin{subfigure}[b]{0.19\textwidth}
    \centering
    \includegraphics[width=\linewidth]{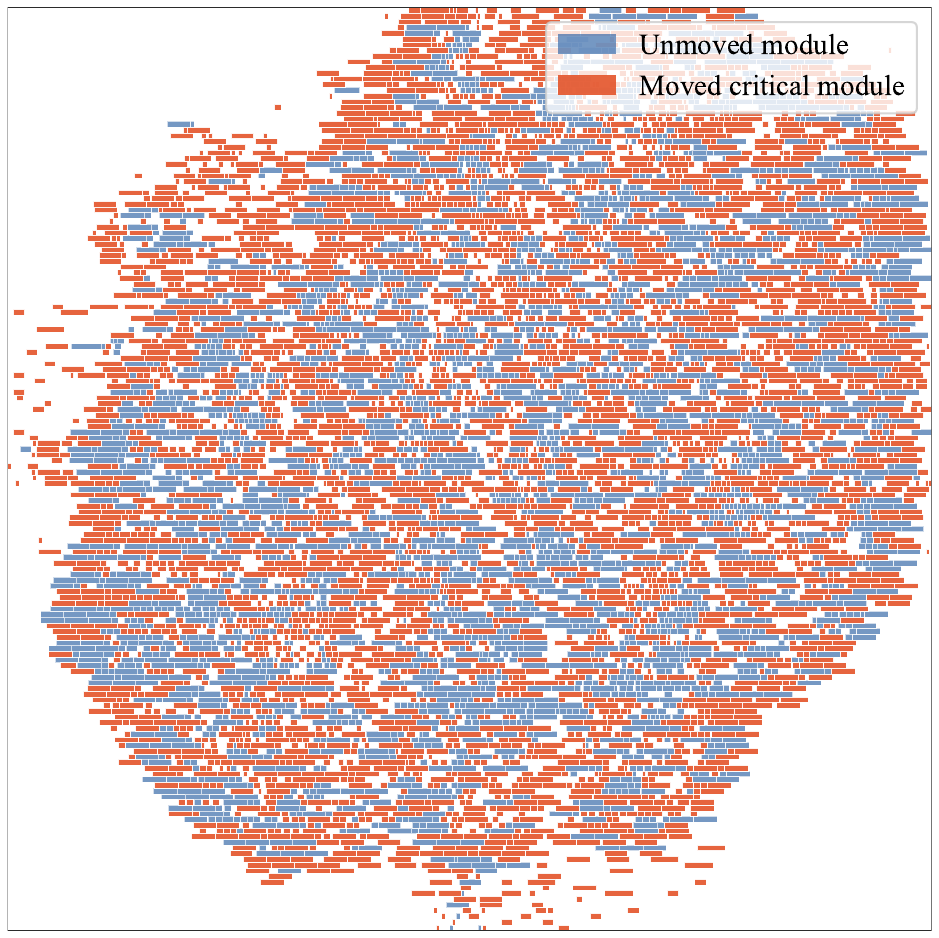}
    \caption{ac97\_top}
    \label{fig:ac97_top_stage3}
\end{subfigure}
\hfill
\begin{subfigure}[b]{0.19\textwidth}
    \centering
    \includegraphics[width=\linewidth]{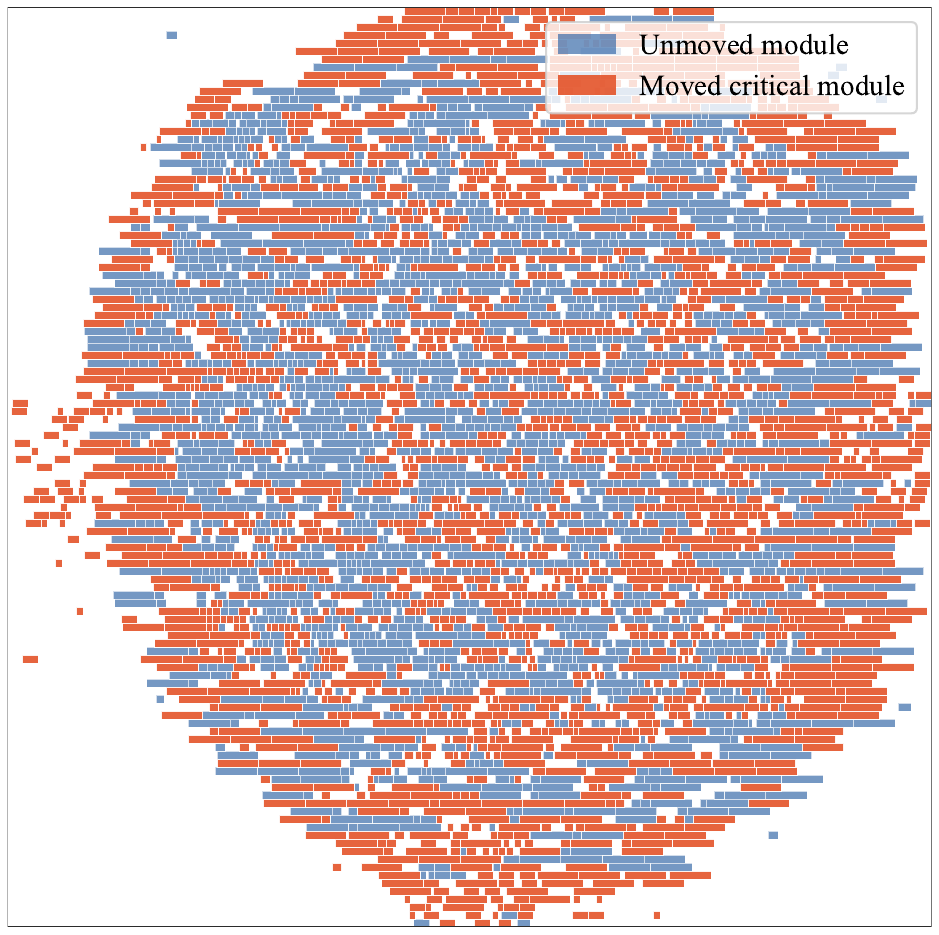}
    \caption{aes}
    \label{fig:aes_stage3}
\end{subfigure}
\hfill
\begin{subfigure}[b]{0.19\textwidth}
    \centering
    \includegraphics[width=\linewidth]{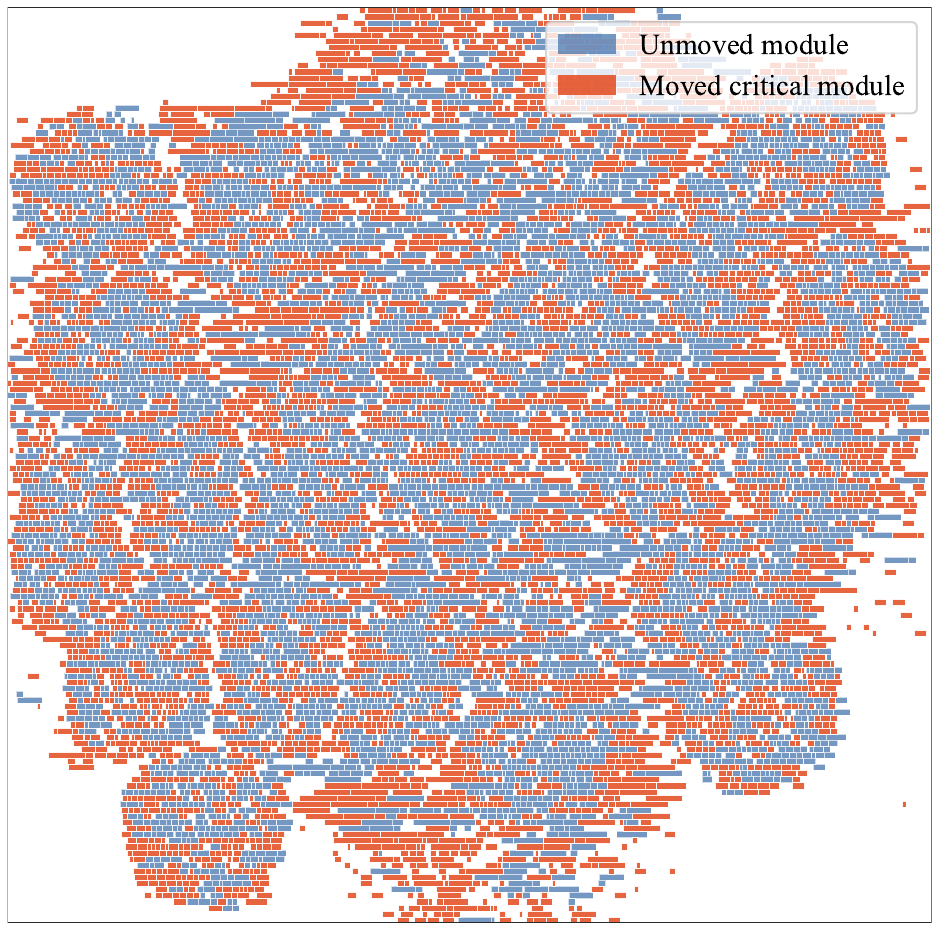}
    \caption{aes\_cipher\_top}
    \label{fig:aes_cipher_top_stage3}
\end{subfigure}
\hfill
\begin{subfigure}[b]{0.19\textwidth}
    \centering
    \includegraphics[width=\linewidth]{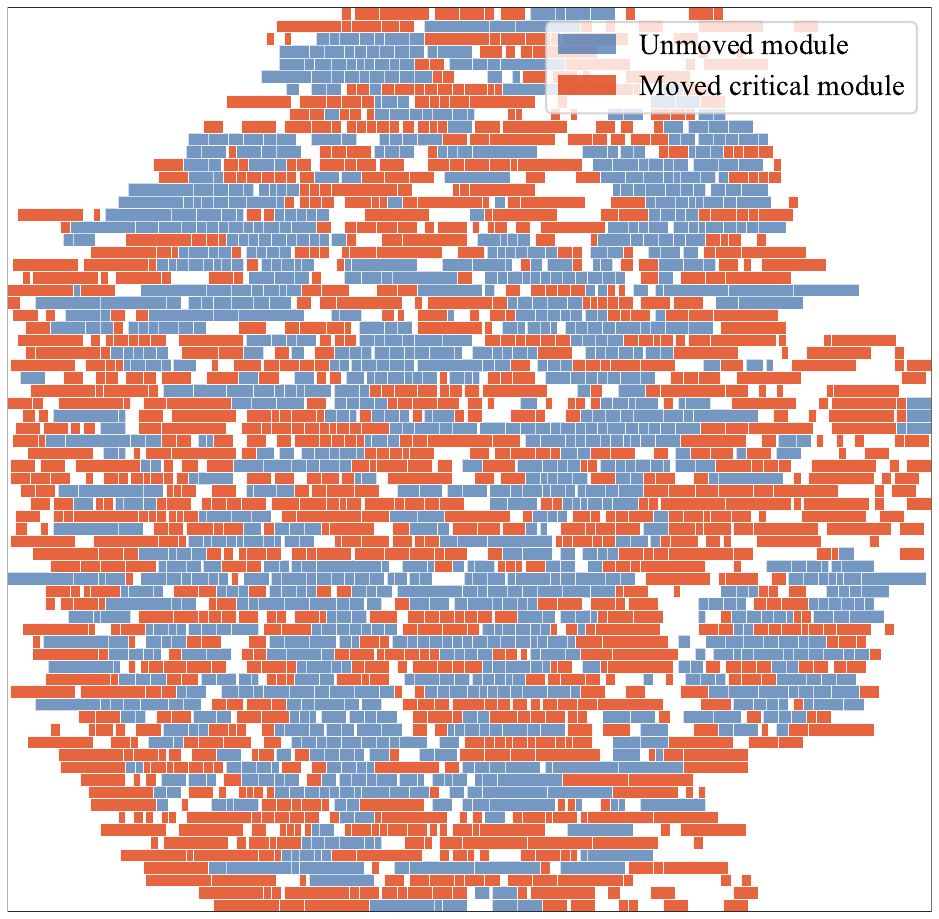}
    \caption{des}
    \label{fig:des_stage3}
\end{subfigure}
\hfill
\begin{subfigure}[b]{0.19\textwidth}
    \centering
    \includegraphics[width=\linewidth]{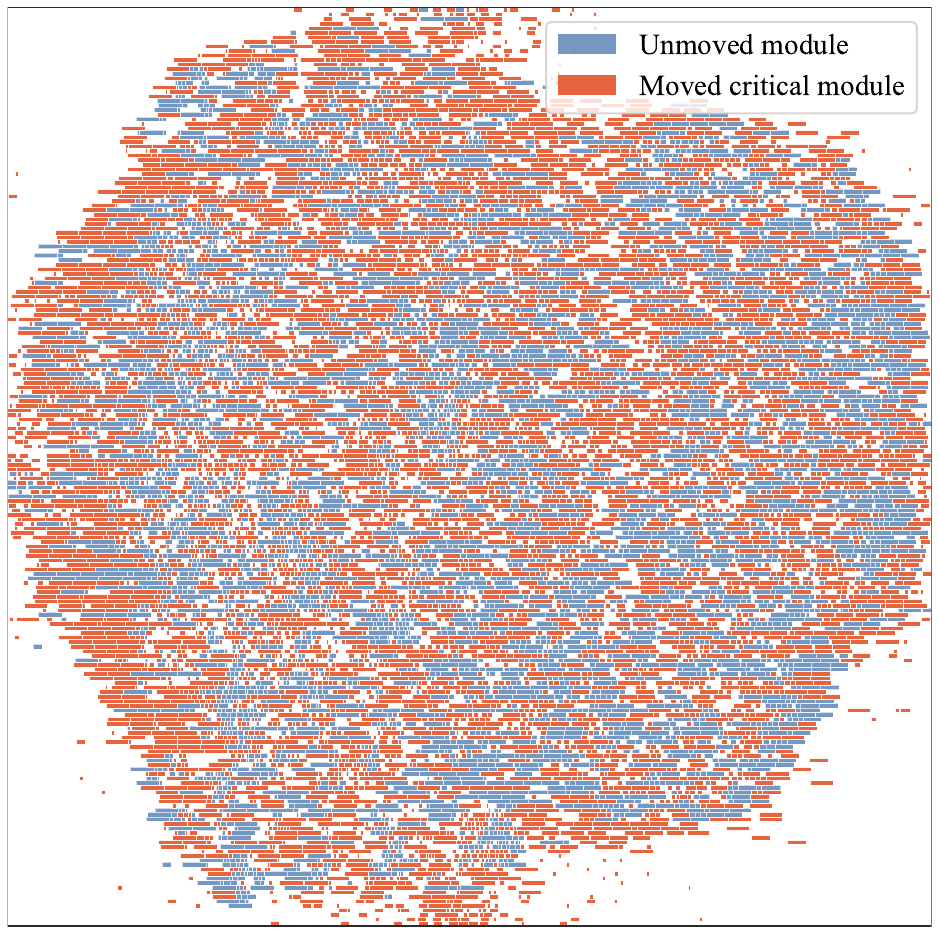}
    \caption{pci\_bridge32}
    \label{fig:pci_bridge32_stage3}
\end{subfigure}

\caption{Placement visualization of five ICCAD2025 benchmark circuits. 
The first and second rows correspond to Stage 2 and Stage 3 optimization, respectively. 
Blue modules remain unchanged, while red indicate tuned critical modules.}
\label{fig:2x5_example}
\end{figure}

Figure~\ref{fig:2x5_example} visualizes the placement refinement process (Stage 3) of C2FPlace on five ICCAD2025 benchmark circuits. It can be observed that a substantial number of critical macros are moved during the refinement process, and these movements are accepted only when they lead to improved solutions. This demonstrates that Stage 3 can effectively further reduce HPWL.

\end{document}